\documentclass[jounal]{IEEEtran}
\usepackage{bm,amsmath,amssymb}
\usepackage{cite}
\ifCLASSINFOpdf
\usepackage[pdftex]{graphicx}
\else
\usepackage[dvips]{graphicx}
\fi
\usepackage{subfigure}

\newtheorem{remark}{Remark}
\usepackage{xcolor}
\usepackage{colortbl,booktabs}
\usepackage{algorithm}
\usepackage{algorithmic}
\usepackage{courier}

\usepackage{multirow}

\begin{document}
	\title{Robust MIMO Detection With Imperfect CSI: A Neural Network Solution}
	
	\author{
				\thanks{This work was supported in part by the National Natural Science Foundation of China under Grant 62271137 and in part by the Southeast University China Mobile Research Institute Joint Innovation Center. The work of Hong Shen was supported by the Natural Science Foundation of Jiangsu Province under Grant BK20201263. The work of Wei Xu was supported in part by the National Natural Science Foundation of China under Grant 62022026 and in part by the Fundamental Research Funds for the Central Universities under Grants 2242022K60002 and 2242023K5003. ({\emph{Corresponding author: Hong Shen}}.)}
		
		%Author 1, Author 2, and Author 3
		Yi~Sun,~\IEEEmembership{Student Member,~IEEE,} Hong~Shen,~\IEEEmembership{Member,~IEEE,} 
		\\Wei~Xu,~\IEEEmembership{Senior Member,~IEEE,} Nan~Hu, and~Chunming~Zhao,~\IEEEmembership{Member,~IEEE}
		
		\thanks{Y. Sun, H. Shen, W. Xu, and C. Zhao are with the National Mobile Communications Research Laboratory, Southeast University, Nanjing 210096, China (e-mail:\{sun\_yi, shhseu, wxu, cmzhao\}@seu.edu.cn).  W. Xu and C. Zhao are also with Purple Mountain
			Laboratories, Nanjing 211111, China. N. Hu is with Institute of Wireless and Terminal Technology, China Mobile Research Institute, Beijing 100000, China (e-mail: hunan@chinamobile.com).} 
			  \thanks{An earlier version of this paper was presented at the 18th International Symposium on Wireless Communication Systems (ISWCS), Hangzhou, China, Oct. 2022 \cite{conf}.}

%		 <-this % stops a space
		
	}

	\maketitle
	
	\begin{abstract}
		In this paper, we investigate the design of statistically robust detectors for multi-input multi-output (MIMO) systems subject to imperfect channel state information (CSI). A robust maximum likelihood (ML) detection problem is formulated by taking into consideration the CSI uncertainties caused by both the channel estimation error and the channel variation. To address the challenging discrete optimization problem, we propose an efficient alternating direction method of multipliers (ADMM)-based algorithm, which only requires calculating closed-form solutions in each iteration. Furthermore, a robust detection network RADMMNet is constructed by unfolding the ADMM iterations and employing both model-driven and data-driven philosophies. Moreover, in order to relieve the computational burden, a low-complexity ADMM-based robust detector is developed using the Gaussian approximation, and the corresponding deep unfolding network LCRADMMNet is further established. On the other hand, we also provide a novel robust data-aided Kalman filter (RDAKF)-based channel tracking method, which can effectively refine the CSI accuracy and improve the performance of the proposed robust detectors. Simulation results validate the significant performance advantages of the proposed robust detection networks over the non-robust detectors with different CSI acquisition methods.  
	\end{abstract}
	
	
		\begin{sloppypar}
	
	\section{Introduction} %semidefinite relaxation (SDR)\cite{SDR} and the
Efficient multi-input multi-out (MIMO) detection algorithms are essential to unleashing the full potential of MIMO techniques, which have been extensively studied for decades in literature \cite{MIMO8944280,50YEAR,LTE}. As is well known, the maximum likelihood (ML) detectors can achieve the optimal performance while with a huge exponentional complexity. Linear detectors, including the zero forcing (ZF) and the linear minimum mean squared error (LMMSE) detectors, enjoy lower complexity but suffer from limited performance. To strike a tradeoff between the performance and  the complexity, some sub-optimal detection algorithms can be applied, such as sphere decoding (SD)\cite{SD}, expectation propagation (EP) \cite{EP}, and approximate message passing (AMP) \cite{AMP}. {The  alternating direction method of multipliers (ADMM) \cite{ADMM8933411} has also been widely used to solve the detection problems from the perspective of convex optimization.} In general, the detection performance highly hinges on the quality of the available channel state information (CSI), which is inevitably imperfect due to some practical issues. However, these classical MIMO detectors are developed based on the assumption of perfect CSI and therefore non-robust to the CSI uncertainties. That means, when we apply these ``mismatched'' detectors by directly regarding the acquired imperfect CSI as if it is perfect, the detection performance can be severely degraded \cite{SNR2005}. To handle this, the CSI imperfection should be considered during the design of a MIMO detector to enhance its robustness. 

%Regarding the robust MIMO detection design, there are typically two classes of approaches corresponding to two different CSI error models, respectively \cite{Statis8944280,maximin8944280,Uncer8944280,Transceiver,Rtransceiver}. One is the stochastic robust design aiming at optimizing the average performance, where the CSI uncertainty is assumed to be a random quantity with its statistical information known by the receiver \cite{Statis8944280}. The other one is the deterministic robust design, which considers the instantaneous CSI uncertainties confined in a bounded region and optimizes the worst-case performance \cite{maximin8944280}. 
In this paper, we focus on the statistically robust MIMO detection with a random CSI error model. {Related works include \cite{Statis8944280,Transceiver,Rtransceiver,WidelyLinear}, where the robust detectors using the linear MMSE criterion and the widely-linear MMSE criterion were studied for the MIMO systems without and with in-phase or quadrature-phase imbalance (IQI), respectively. Alternatively, a linear minimum error probability (MEP) detector with a given length of pilots was derived to enable ultra-reliable and low latency communication (URLLC) under imperfect CSI \cite{MEP}.}  {Furthermore, in \cite{Model8509622}, the authors investigated a robust orthogonal approximate mesasge passing (OAMP) algorithm as well as its corresponding network by exploiting the statistics of imperfect CSI, which can exhibit stronger robustness against CSI errors than the conventional OAMP algorithm \cite{OAMPNet}.} The optimal robust ML detectors were discussed in \cite{Any2005,Data7894280,QAM86,reduce} by incorporating the statistical distribution of channel estimation errors. Although their performance advantages have been validated by both theoretical analysis and simulation results in \cite{Any2005}, the exponential complexity impedes their practicality.  On the other hand, in most of these works, the CSI errors were characterized by some relatively simple models, e.g., the independent and identically distributed (i.i.d.) Gaussian model, which, however, is inconsistent with practical systems with spatial correlation. Therefore, a more realistic CSI error model needs to be adopted for the robust MIMO detection design.   

Recently, the great advancement of deep learning (DL)  has motivated the researches on the DL-aided wireless communications \cite{Deep8663966,Model5338,unroll,survey,xw2}. In particular, DL has been successfully applied in the physical layer communication techniques, such as channel estimation \cite{chanest,ICINet}, MIMO precoding \cite{Precoder,linqi}, and, as our main interest, MIMO detection \cite{DetNet52521,LAN8715338,LCG,Binary,Low1312,xw1}, {which have been thoroughly reviewed in \cite{survey}.} %Thanks to the powerful learning ability based on the training data, DL has the potential to outperform existing schemes especially under certain challenging scenarios. On the other hand, DL can also be used to  reduce the complexity of the traditional algorithms. 
The DL-based designs can be categorized into data-driven and model-driven types. The data-driven DL regards the mapping function from the input to the desired output as a black box and directly learns the function by training a network of a specific structure \cite{power}, whose performance highly relies on the training dataset. Consequently, the overfitting problem may occur with an insufficient dataset or a large amount of trainable parameters. To relax the requirement on the training data, the model-driven DL that leverages the expert knowledge has emerged as an alternative \cite{Model5338}. One of the most popular model-driven DL approaches is ``deep unfolding'', which unfolds an existing iterative algorithm into network layers with trainable parameters \cite{unroll}. Accordingly, the inherent mechanism of the original algorithm can be maintained and only several parameters need to be learned. The deep unfolding based MIMO detection was first investigated in \cite{DetNet52521}, where a network called DetNet was built by mimicking the projected gradient algorithm for the ML optimization problem. Inspired by this, a variety of MIMO detection networks, such as OAMPNet \cite{OAMPNet}, ADMMNet \cite{LAN8715338}, and LcgNet \cite{LCG}, were proposed based on the idea of deep unfolding. However, these networks are designed based on the availability of perfect CSI at the receiver,  which can suffer from performance degradation with mismatched CSI.

Motivated by the advantage of the deep unfolding technique, in this paper, we advocate two model-driven robust MIMO detection networks, called RADMMNet and LCRADMMNet, to combat CSI imperfection. {Specifically, inspired by the attractive performances of ADMM-based MIMO detectors under perfect CSI \cite{ADMIN,Efficient75673}, %such as ADMIN \cite{ADMIN} and PS-ADMM \cite{Efficient75673}, 
we first apply the ADMM framework to derive the solutions to two robust ML detection problems, based on which RADMMNet and LCRADMMNet are further developed. The performance superiorities of the proposed networks over conventional non-robust MIMO detectors are validated via numerical simulations.} The main contributions of this paper include:
\begin{itemize}
	\item Focusing on a spectrum-efficient frame structure with pilots only used in the first block, we model the corresponding CSI imperfection by including both the channel estimation error and the channel variation, based on which a robust ML detection problem is formulated.
	\item {In order to address the complicated robust ML detection problem, we develop an ADMM-based algorithm that only involves the calculations of closed-form expressions in each ADMM iteration, which, to the best of our knowledge, has not been investigated in previous works. Furthermore, by unfolding the ADMM iterations with non-trivial simplifications, a robust detection network, termed as RADMMNet, is established to learn the layer-wise parameters instead of performing an exhaustive search. The network design is further enhanced by transferring intermediate variables and incorporating trainable convolutional layers, so as to enjoy the advantages of both the model-driven deep unfolding technique and the introduced data-driven structure.}
	\item By exploiting a Gaussian approximation for the CSI error, we obtain a simplified reformulation of the robust ML detection problem. {Based on this formulation, an ADMM-based robust detector is proposed to approach the performance of our first detector with lower complexity,  where closed-form solutions are also derived in each ADMM iteration.} The corresponding deep unfolding network, termed as LCRADMMNet is further constructed in a similar way as RADMMNet.
	\item A robust data-aided Kalman filter-based channel tracking method is established to mitigate the error propagation caused by channel aging, where the pre-estimated data symbols are utilized to improve the accuracy of the channel estimation, thereby enhancing the robust detection performance.
\end{itemize}

The rest of this paper is organized as follows. In Section II, we describe the system model and formulate the robust ML detection problem. Section III and Section IV elaborate the design of a robust detector and a low-complexity robust detector, respectively, wherein their corresponding deep unfolding networks are also developed. Section V presents a robust data-aided Kalman filter based channel tracking method. Numerical results are provided in Section VI, followed by the conclusion drawn in Section VII. 

%{{\emph{Notations}}: Throughout the paper, italic lowercase letters, boldface lowercase letters, and boldface uppercase letters are used to represent scalars, vectors, and matrices, respectively. %$\rm{diag}\left(\bf a\right)$ stands for a diagonal matrix with $\bf a$ being the diagonal vector. 
%For each matrix $\bf A$, ${\bf A}^T$, ${\bf A}^H$, $\rm{det}\left(\bf A\right)$ and $\rm{tr}\left(\bf A\right)$ denote its transpose, conjugate transpose, determinant, and trace, respectively. $\rm{vec}\left(\bf A\right)$ represents the vectorized form of $\bf A$ by stacking all the column vectors, and ${\bf A}\left({p_1}:{p_2},{q_1}:{q_2}\right)$ returns the submatrix of $\bf A$ comprised of its entries from row $p_1$ to row $p_2$ and from column $q_1$ to column $q_2$. $\bf A \otimes \bf B$ and $\bf A \odot \bf B$ are the Kronecker product and the Hadamard product of $\bf A$ and $\bf B$, respectively. In addition, ${\bf I}_k$ denotes the identity matrix of size $k \times k$, ${\bf{1}}_{{k_1} \times {k_2}}$ denotes the all-ones vector or matrix of size ${{k_1} \times {k_2}}$, and ${\bf 0}_{{m_1} \times {m_2}}^{\backslash i,j}$ denotes the all-zeros matrix of size ${{m_1} \times {m_2}}$ except the $(i,j)$-th entry being $1$. Finally, ${\rm Re}\{\cdot\}$ and ${\rm Im}\{\cdot\}$ extract the real and imaginary parts of the input, respectively.} %${\mathbb E}\{\cdot\}$ takes the expectation of the input, and $\left\| \cdot \right\|_2$ is the ${\mathcal{L}}_2$-norm of the input vector.}

	\section{System Model and Problem Formulation}
	\subsection{System Model}

\begin{figure}[t]
	\centering
	\includegraphics[width=8cm]{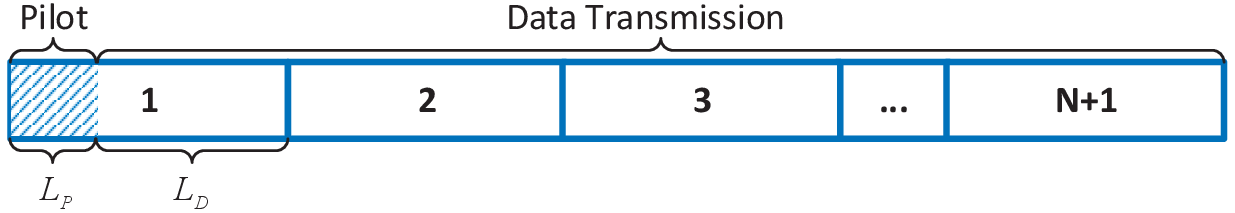}
	\caption{Frame structure.}
	\label{fig1}
\end{figure}

Consider an uplink MIMO system, where a base station (BS) equipped with $M$ receive antennas serves $K$ single-antenna users. The channel remains constant during a coherent block comprised of $L$ symbols and varies across blocks. As shown in Fig. \ref{fig1}, we adopt the frame structure that contains $N+1$ coherence blocks to save the pilot overhead. Specifically, in each frame, ${L_P}$ pilot symbols followed by ${L_D}$ data symbols are sent in the first block, while the remaining $N$ blocks are only used for data transmission. Note that, compared to the conventional frame structure that inserts pilots in each coherence block \cite{DACE}, the pilot overhead reduces from $\frac{{L_P}}{L}$ to $\frac{{L_P}}{(N+1)L}$, which is thus much more spectrum-efficient. For a convenient reference, the main symbols used in this paper are summarized in Table I at the top of the next page.

\begin{table*}[]
	\caption{Main Symbols Used in This Paper}
	\resizebox{\textwidth}{!}{
	\resizebox{\columnwidth}{!}{%
		\begin{tabular}{|c|c|}
			\hline
			{Symbol}                                                                & {Definition}                                                                                                                          \\ \hline
			{$M$,$K$}                                                                   & {the numbers of receive antennas at the BS and single-antenna users}                                                                                            \\ \hline
%			{$K$}                                                                   & {the number of single-antenna users}                                                                                                  \\ \hline
%			{$L$}                                                                   & {the number of symbols in a coherent block}                                                                                           \\ \hline
			{$L$,$L_P$,$L_D$}                                                           & {the numbers of symbols in a coherent block, pilot symbols  in the first block, and data symbols in the first block}                                                                      \\ \hline
			{$N$}                                                                   & {the number of data blocks for data transmission}                                                                                      \\ \hline
			%						${\bf S}_P$                                                           & the pilot symbol matrix                                                                                                             \\ \hline
			{$\rho_k$}                                                              & {the channel temporal correlation coefficient of the $k$-th user}                                                                     \\ \hline
%			${\bf C}_{\bf h}$                                                     & the covariance of the channel vector                                                                                                \\ \hline
			{${\bf x}$, ${\bf H}$, ${\bf h}$, ${\bf C}_{\bf h}$}                                             & {the transmitted vector, the channel matrix, the channel vector, and the channel covariance}                                                         \\ \hline
			{$\hat{\bf h}$, ${\bm \Delta}{\bf h}$, ${\bf \Sigma}_{\bf h}$} & {the channel estimate vector, the CSI error vector, and the corresponding error covariance}                        \\ \hline
			{$\hat{\bf H}$, ${\bm \Delta}{\bf H}$, ${\bf \Sigma}_{\bf H}$} & {the channel estimate matrix, the CSI error matrix, and the corresponding error covariance}                        \\ \hline
				{$\hat{\bf x}$, ${\bm \Delta}{\bf x}$, ${\bf \Sigma}_{\bf x}$} & {the data estimate vector, the data estimation error vector, and the corresponding error covariance}                        \\ \hline
%			{$[n \left|{n - 1}\right.]$}                                            & {the \emph{a prior} quantity of the $n$-th block based on the previous information of the $(n-1)$-th block}        \\ \hline
%			{$[n \left|{n}\right.]$}                                                & {the \emph{a posterior} quantity of the $n$-th block obtained by uitilizing the observation of the $n$-th block} \\ \hline
		\end{tabular}%
	}}
\end{table*}

To characterize the channel aging effect, we use the first-order Gauss-Markov process as in \cite{Markov,9445013}. Define ${\bm{\Lambda}} \triangleq {\rm{diag}}\left({\rho _1},{\rho _2}, \cdots ,{\rho _K}\right) \otimes {{\bf{I}}_M}$ and ${\bar{\bm{\Lambda}}} \triangleq {\rm{diag}}\left(\sqrt {1 - \rho _1^2} ,\sqrt {1 - \rho _2^2} , \cdots ,\sqrt {1 - \rho _K^2} \right) \otimes {{\bf{I}}_M}$, where ${\rho}_k$ is the  channel temporal correlation coefficient of the $k$-th user related to the channel aging speed. %Generally, ${\rho}_k$ can be determined according to the widely-used Jakes' model, i.e., ${\rho _k} = {J_0}\left(2\pi {v_k}{f_c}{T_s}/c\right)$, where ${J_0}\left(\cdot\right)$ is the zero-th order Bessel function of the first kind, $v_k$ is the moving speed of the $k$-th user, ${f_c}$ is the carrier frequency, $T_s$ is the sampling period, and $c$ is the speed of light. 
Then, the channel variation between two neighboring blocks can be represented
\begin{equation}
\begin{aligned} 
{\bf{h}}[n] = {\bm{\Lambda}}{\bf{h}}[n - 1] + {\bar{\bm{\Lambda}}}{\bf w}[n], n = 2,3,\cdots,N+1,
\end{aligned}
\end{equation}
where ${{\bf{h}}[n]}\in \mathbb{C}^{MK \times 1}$ denotes the vectorized form of the channel matrix ${{\bf{H}}[n]}\in \mathbb{C}^{M \times K}$ at the $n$-th block and follows a Gaussian distribution with zero mean and covariance ${{\bf{C}}_{\bf{h}}}$, and ${\bf{w}}[n] \in \mathbb{C}^{MK \times 1}$ denotes the zero-mean spatially correlated innovation process with covariance ${{\bf{C}}_{\bf{h}}}$. Furthermore, it can be easily inferred that
%\vspace{-1mm}
\begin{equation} 
\begin{aligned}
	{{\bf{h}}[n]} = {{\bm{\Lambda}}}^{n-1}{{\bf{h}}[1]} +  \tilde{\bf{w}}[n], n = 2,3,\cdots,N+1,
\end{aligned}
\end{equation}
where ${{\bf{h}}[1]}$ represents the channel vector of the first block, and the channel error ${{\bf{\tilde w}}}[n]$ follows a Gaussian distribution with zero mean and covariance ${{\bf{C}}_{\tilde{\bf{w}}}}[n]$, which is given by
 	\begin{equation} 
% 		\begin{small}
 			\begin{aligned}
 		{{\bf{C}}_{\tilde{\bf{w}}}}[n]= \sum\limits_{n' = 1}^{n-1} {{{\bm{\Lambda}}^{n'-1}}{\bar{\bm{\Lambda}}}{{\bf{C}}_{\bf{h}}}{\bar{\bm{\Lambda}}}{{\bm{\Lambda}}^{n'-1}}}, n = 2,3,\cdots,N+1.
\end{aligned}
%	\end{small}
%\end{small}
 \end{equation}
For the first block, letting ${{\bf{S}}_P} \in {\mathbb{C}^{K \times {L_p}}}$ denote the pilot sequences, the received pilot signal can be expressed as
%\vspace{-4mm}
	\begin{equation}
	{{\bf{Y}}_P}={{\bf{H}}[1]}{{\bf{S}}_P}{\bf{ + }}{{\bf{Z}}_P},
\end{equation}
{where ${{\bf{Z}}_P}\in \mathbb{C}^{M \times {L_p}}$ is the noise matrix whose elements are independently identically distributed (i.i.d.) Gaussian variables with zero mean and variance ${\sigma}^2$. Vectorizing (4) yields}
	\begin{equation}
	{{\bf{y}}_{P}} = {\bf{P}}{{\bf{h}}[1]} + {{\bf{z}}_{P}},
\end{equation}
where ${{\bf{y}}_{P}} = {\rm{vec}}({{\bf{Y}}_P})$, ${\bf{P}} = {\bf{S}}_P^T \otimes {{\bf{I}}_M}$, and ${{\bf{z}}_{P}} = {\rm{vec}}({{\bf{Z}}_P})$. {Considering the commonly used orthogonal pilots with the normalized average power of each pilot symbol, i.e., ${{\bf{S}}_P}{\bf{S}}_P^H = {L_P}{{\bf{I}}_K}$, it then follows that ${{\bf{P}}^H}{\bf{P}} = {L_P}{{\bf{I}}_{MK}}$.} From (2), (3), and (5), the linear MMSE (LMMSE) channel estimate of the $n$-th block with given ${{\bf{y}}_{P}}$ can be obtained via \cite{Markov}
	\begin{equation}
		\begin{aligned}
{{\hat{\bf h}}[n]} = {{\bm{\Lambda}}}^{n-1}{{\bf{C}}_{\bf h}}&{\left({L_P}{{\bf{C}}_{\bf h}} + {\sigma ^2}{{\bf{I}}_{MK}}\right)^{ - 1}}{{\bf{P}}^H}{{\bf{y}}_P}, \\
		&\quad\quad\quad\quad\quad\quad\quad n = 1,2,\cdots,N+1.
\end{aligned}
\end{equation}
Accordingly, we define the CSI uncertainty as ${{\bm \Delta}{\bf h}[n]} = {{\bf{h}}[n]} - {{\hat{\bf h}}[n]}$, which is Gaussian distributed with zero mean and the covariance written as
	\begin{equation}
			\begin{small}
		\begin{aligned}
			{ {{\bm{\Sigma }}_{\bf{h}}}[n] = \left\{ {\begin{array}{*{20}{c}}
					{{\sigma ^2}{{\bf{C}}_{\bf{h}}}{{\left( { {L_P} {{\bf{C}}_{\bf{h}}} + {\sigma ^2}{{\bf{I}}_{MK}}} \right)}^{ - 1}},\quad \quad n = 1},\\
					{{\sigma ^2}{{\bf{\Lambda }}^{n - 1}}{{\bf{C}}_{\bf{h}}}{{\left( { {L_P} {{\bf{C}}_{\bf{h}}} + {\sigma ^2}{{\bf{I}}_{MK}}} \right)}^{ - 1}}{{\bf{\Lambda }}^{n - 1}} + {{\bf{C}}_{{\tilde{\bf w}}}}[n],}\\{\quad \quad \quad \quad \quad \quad \quad \quad \quad \quad \quad \quad n = 2,3, \cdots ,N + 1.}
			\end{array}} \right.}
\end{aligned}
\end{small}
\end{equation}
\subsection{Problem Formulation}
We now consider the data transmission. Denote ${{\bf{x}}_t[n]\in \mathbb{C}^{K \times 1}}$ as the transmitted vector at the $t$-th time slot of the $n$-th block, each entry of which is drawn from a quadrature amplitude modulation (QAM) constellation set $\cal{X}$. Then, the received signal is given by
\begin{equation} 
	\begin{aligned}
		{{\bf{y}}_{t}[n]}& = {{\bf{H}}[n]}{{\bf{x}}_t[n]} + {{\bf{z}}_{t}[n]}, \\&\quad t = 1,2,\cdots,{L},\quad  n = 1,2,\cdots,N+1,
	\end{aligned}
\end{equation}
where the noise vector ${{\bf{z}}_{t}[n]} \sim {\cal{CN}}\left({\bf{0}},{{\sigma}^2} {\bf{I}}_{M}\right)$. For simplicity of the notation, the indices $t$ and $n$ will be omitted if there is no confusion. 

With the imperfect CSI in (6) available, one can recover the transmit signals by directly using the channel estimate as if it is perfect, which corresponds to the mismatched ML criterion:
\begin{equation}
	{\bf{\hat x}} = \mathop {\arg \min }\limits_{{\bf{x}} \in {{\cal{X}}^K}} \left\| {{\bf{y}} - {\bf{\hat Hx}}} \right\|_2^2,
\end{equation}
where ${{\hat{\bf H}}}$ is the matrix form of ${{\hat{\bf h}}}$. However, this is not the optimal criterion since the CSI uncertainty is neglected. To fill this gap, we first recast (8) by
\begin{equation} 
	{{\bf{y}}} = {{\bf{X}}}{{\bf{h}}} + {{\bf{z}}},
\end{equation}
where ${\bf{X}} = {\bf{x}}^T \otimes {{\bf{I}}_M}$. Then, by regarding ${{\bf{h}}}$ as a conditional Gaussian vector with conditional mean ${\bf{\hat h}}$ and covariance ${{\bm{\Sigma }}_{\bf{h}}}$ and performing some mathematical manipulations as in \cite{Data7894280}, the robust ML function can be derived as 
\begin{equation} 
p\left({\bf{y}}\left| {{\bf{X}},{\bf{\hat h}}} \right.\right) = C \det\left({{\bf{R}}_{\bf{X}}^{ - 1}}\right)\exp \left({\bf{q}}_{\bf{X}}^H{\bf{R}}_{\bf{X}}^{ - 1}{{\bf{q}}_{\bf{X}}}\right),
\end{equation}
where $C$ is a constant, ${{\bf{q}}_{\bf{X}}} = \frac{1}{{{\sigma ^2}}}{{\bf{X}}^H}{\bf{y}} + {\bm{\Sigma }}_{\bf{h}}^{ - 1}{\bf{\hat h}}$, ${{\bf{R}}_{\bf{X}}} = \frac{1}{{{\sigma ^2}}}{{\bf{X}}^H}{\bf{X}} + {\bm{\Sigma }}_{\bf{h}}^{ - 1}$, and ${\bm{\Sigma }}_{\bf{h}}$ is given in (7). Furthermore, maximizing (11) yields the subsequent robust ML detection criterion:
\begin{equation}
	{\bf{\hat x}} = \mathop {\arg \min }\limits_{{\bf{x}} \in {{\cal{X}}^K}} \ln \det \left({{\bf{R}}_{\bf{X}}}\right) - {\bf{q}}_{\bf{X}}^H{\bf{R}}_{\bf{X}}^{ - 1}{{\bf{q}}_{\bf{X}}}.
\end{equation}
Generally, due to the $K$-dimensional discrete constraint ${\bf{x}} \in {{\cal{X}}^K}$, the global optimal solution of the above problem can only be acquired via an exhaustive search, which requires an unacceptable exponential complexity. In the following section, we provide a neural network based solution to the problem with excellent performance and much reduced complexity.

\section{Robust ADMM Detector Based Network Design}
In this section, we first devise an ADMM-based algorithm to address problem (12). Based on this, we then provide the design of the proposed robust detection network RADMMNet.
\subsection{Robust ADMM Detector}
In order to develop an efficient robust detector, we first reformulate problem (12) into a tractable form. It is known from \cite{Efficient75673} that the real and the imaginary parts of a $4^Q$-QAM signal can be decomposed into a weighted sum of $Q$ binary variables, respectively. Using this representation, problem (12) can be equivalently expressed by
\begin{equation} 
	\begin{aligned}
		\mathop {\min }\limits_{{\bf{x}},{\{{{\bf{v}}_q},\forall q\}} } &\ln \det \left({{\bf{R}}_{\bf{X}}}\right) - {\bf{q}}_{\bf{X}}^H{\bf{R}}_{\bf{X}}^{ - 1}{{\bf{q}}_{\bf{X}}}
		\\ {\text{s.t.}} \quad &{\bf{x}} - \frac{1}{{{\alpha _Q}}}\sum\limits_{q = 1}^Q {{2^{q - 1}}{{\bf{v}}_q}}  = {\bf{0}}, \\\quad& {\mathop{\rm Re}\nolimits} \{ {{\bf{v}}_q}\} ,{\mathop{\rm Im}\nolimits} \{ {{\bf{v}}_q}\}  \in {\{  - 1,1\} ^K},q = 1, \cdots ,Q,
	\end{aligned}
%\end{small}
\end{equation}
where $\alpha_Q $ is the power normalization factor of QAM signals and ${\{  - 1,1\} ^K}$ denotes the binary set of $K \times 1$ vectors with each entry being $1$ or $-1$. Furthermore, we relax the discrete binary constraints by the boxed constraints and then impose a sum of quadratic penalty terms to the objective, yielding the following problem: 
\begin{equation} 
	\begin{aligned}
		\mathop {\min }\limits_{{\bf{x}},{\{ {{\bf{v}}_q},\forall q\}} } &\ln \det \left({{\bf{R}}_{\bf{X}}}\right) - {\bf{q}}_{\bf{X}}^H{\bf{R}}_{\bf{X}}^{ - 1}{{\bf{q}}_{\bf{X}}} - \sum\limits_{q = 1}^Q {{\beta _q}\left\| {{{\bf{v}}_q}} \right\|_2^2}
		\\{\rm{s.t.}}\quad &{\bf{x}} - \frac{1}{{{\alpha _Q}}}\sum\limits_{q = 1}^Q {{2^{q - 1}}{{\bf{v}}_q}}  = {\bf{0}}, \\\quad & {\mathop{\rm Re}\nolimits} \{ {{\bf{v}}_q}\} ,{\mathop{\rm Im}\nolimits} \{ {{\bf{v}}_q}\}  \in {[- 1,1] ^K},q = 1, \cdots ,Q,
	\end{aligned}
%\end{small}
\end{equation}
where ${\beta_q}>0,q = 1, \cdots ,Q $ are the penalty parameters, and ${[ - 1,1] ^K}$ denotes the set of $K \times 1$ vectors with each entry lying in the interval $[-1,1]$. Note that the added penalty term $-\sum\limits_{q = 1}^Q {{\beta _q}\left\| {{{\bf{v}}_q}} \right\|_2^2}$ makes the integer solutions more preferable  and therefore tightens the boxed constraints \cite{Binary}. Based on the reformulation, we apply the ADMM algorithm to tackle the problem as follows \cite{ADMM8933411}.      

To facilitate the application of the ADMM framework,  we construct the scaled augmented Lagrangian function of problem (14) as
\begin{equation}
%	\begin{small}
	\begin{aligned}
	{L_\mu }\left({\bf{x}},\{ {{\bf{v}}_q}\}_{q = 1}^Q ,{\bm{\lambda }}\right) = \ln \det ({{\bf{R}}_{\bf{X}}}) - {\bf{q}}_{\bf{X}}^H{\bf{R}}_{\bf{X}}^{ - 1}{{\bf{q}}_{\bf{X}}} \\- \sum\limits_{q = 1}^Q {{\beta _q}\left\| {{{\bf{v}}_q}} \right\|_2^2}   + \frac{\mu }{2}\left\| {{\bf{x}} - \frac{1}{{{\alpha _Q}}}\sum\limits_{q = 1}^Q {{2^{q - 1}}{{\bf{v}}_q}}  + {\bm{\lambda }}} \right\|_2^2,
\end{aligned}
%\end{small}
\end{equation}
where ${\bm{\lambda }}$ and $\mu>0$ denote the scaled dual variables and the corresponding penalty parameter, respectively. Thus, the ADMM procedure can be described as follows:
\begin{subequations}
%	\begin{small}
	\begin{equation}
%			\begin{small}
			\begin{aligned}
		\{ {{\bf{v}}_q^i}\} _{q = 1}^Q = \mathop {\arg \min }\limits_{{{\bf{v}}_q} \in {{\mathbb B}^K}}  {L_\mu }\left({\bf{x}}^{i-1},\{ {{\bf{v}}_q}\}_{q = 1}^Q ,{\bm{\lambda }^{i-1}}\right), %- \sum\limits_{q = 1}^Q {{\beta _q}\left\| {{{\bf{v}}_q}} \right\|_2^2}  + \frac{\mu }{2}\left\| {{\bf{x}}^{i-1} - \frac{1}{{{\alpha _Q}}}\sum\limits_{q = 1}^Q {{2^{q - 1}}{{\bf{v}}_q}}  + {\bm{\lambda }^{i-1}}} \right\|_2^2,
	\end{aligned}
%\end{small}
	\end{equation}
	\begin{equation}
%					\begin{small}
			\begin{aligned}
		{{\bf{x}}^i} = \mathop {\arg \min }\limits_{\bf{x}}  {L_\mu }\left({\bf{x}},\{ {{\bf{v}}^{i}_q}\}_{q = 1}^Q ,{\bm{\lambda }^{i-1}}\right),% \ln \det ({{\bf{R}}_{\bf{X}}}) - {\bf{q}}_{\bf{X}}^H{\bf{R}}_{\bf{X}}^{ - 1}{{\bf{q}}_{\bf{X}}}+ \frac{\mu }{2}\left\| {{\bf{x}} - \frac{1}{{{\alpha _Q}}}\sum\limits_{q = 1}^Q {{2^{q - 1}}{{\bf{v}}_q^i}}  + {\bm{\lambda }}^{i-1}} \right\|_2^2,
		%{{\bf{x}}^i} = \mathop {\arg \min }\limits_{\bf{x}} {L_\mu }({\bf{x}},\{ {\bf{v}}_q^i\} _{q = 1}^Q,{{\bm{\lambda }}^{i - 1}}),
	\end{aligned}
%\end{small}
	\end{equation}
	\begin{equation}
%		\begin{small}
			\begin{aligned}
		{{\bm{\lambda }}^i} = {{\bm{\lambda }}^{i - 1}} + {{\bf{x}}^i} - \frac{1}{{{\alpha _Q}}}\sum\limits_{q = 1}^Q {{2^{q - 1}}{\bf{v}}_q^i},
	\end{aligned}
%\end{small}
	\end{equation}
\end{subequations}
where the superscript $i$ denotes the index of ADMM iterations and ${\mathbb B}^K$ denotes the set of $K \times 1$ vectors, both the real and the imaginary parts of whose entries belong to the interval $[-1,1]$.

For subproblem (16a), the difficulty of obtaining the optimal solution lies in the fact that $\{ {{\bf{v}}_q}\} _{q = 1}^Q$ are coupled with each other. Nevertheless, we find that ${L_\mu }({{\bf{x}}^{i - 1}},\{ {{\bf{v}}_q}\} _{q = 1}^Q,{{\bm{\lambda }}^{i - 1}})$ is convex with respect to (w.r.t.) each ${{\bf{v}}_q}$ when the condition ${\mu {4^{q - 1}} - 2\alpha _Q^2{\beta _q}}>0$ is satisfied, which inspires us to apply the block coordinate descent (BCD) method \cite{zhangyu}. Consequently, by taking the derivative of (15) w.r.t. each ${{\bf{v}}_q}$ and setting it to zero, a closed-form solution to ${\bf{v}}_q$ can be achieved by
	\begin{equation}
	{\bf{v}}_q^{i} = \mathop {\cal P}\nolimits_{{{\mathbb{B}}^K}} \left\{\frac{{{2^{q - 1}}}{\mu}{{{{\bm{\eta}}}_q^{i}}}}{{\mu {4^{q - 1}} - 2\alpha _Q^2{\beta _q}}} \right\}, \quad q = 1,2, \cdots ,Q,
\end{equation}
where  $ {{{\bm{\eta}}}_q^{i}} = { {\alpha _Q}\left({{{\bf{x}}}^{i-1}} + {{{\bm{\lambda }}}^{i - 1}}\right) -  \sum\limits_{p < q} {{2^{p - 1}}{{\bf{v}}}_p^{i}}  - \sum\limits_{p > q} {{2^{p - 1}}{{\bf{v}}}_p^{i-1}} }$ and $\mathop {\cal P}\nolimits_{{\mathbb{B}}^K}\{\cdot\}$ means projecting the real and the imaginary parts of each entry of the input vector onto the interval $[-1,1]$.%, respectively. %Note that (17) is the optimal solution to subproblem (16a) for the quadrature phase shift keying (QPSK) case, i.e., $Q = 1$.

The remaining subproblem (16b) is more challenging due to its complicated non-convex objective. To make it tractable, we replace the non-trivial function $\ln \det \left({{\bf{R}}_{\bf{X}}}\right) - {\bf{q}}_{\bf{X}}^H{\bf{R}}_{\bf{X}}^{ - 1}{{\bf{q}}_{\bf{X}}}$ with an appropriate upperbound surrogate function in each ADMM iteration. In this way, the original minimization problem is reduced to minimizing the surrogate function that is easier to handle. Concretely, we first establish an upper bound to the function $\ln \det\left({{\bf{R}}_{\bf{X}}}\right)$. Since the function is concave in ${{\bf{R}}_{\bf{X}}}$, it is majorized by its first-order Taylor expansion as
\begin{equation} 
%\begin{small}
	\begin{aligned}
		\ln \det \left({{\bf{R}}_{\bf{X}}}\right) &\leq \ln \det \left({{\bf{R}}_{{{\bf{X}}^{i-1}}}}\right) + {\rm{tr}}\left({\bf{R}}_{{{\bf{X}}^{i-1}}}^{ - 1}\left({{\bf{R}}_{\bf{X}}} - {{\bf{R}}_{{{\bf{X}}^{i-1}}}}\right)\right) \\
		%&= tr\left( {{\bf{R}}_{{{\bf{X}}^{i}}}^{ - 1}\left(\frac{1}{{{\sigma ^2}}}{{\bf{X}}^H}{\bf{X}} + {\bm{\Sigma }}_{\bf{h}}^{ - 1})} ) + {\rm{constant}} \\
		&\mathop  = \limits^{(a)}  \frac{1}{{{\sigma ^2}}}{\rm{tr}}\left( {{\bf{R}}_{{{\bf{X}}^{i-1}}}^{ - 1}{{\bf{X}}^H}{\bf{X}}} \right) + {\rm{constant}}\\
		&\mathop  = \limits^{(b)}  {{\bf{x}}^H}{{\bf{C}}^{i-1}}{\bf{x}} + {\rm{constant}},
	\end{aligned}
%\end{small}
\end{equation}
where (a) uses the definition of ${{\bf{R}}_{\bf{X}}} = \frac{1}{{{\sigma ^2}}}{{\bf{X}}^H}{\bf{X}} + {\bm{\Sigma }}_{\bf{h}}^{ - 1}$, and (b) is acquired based on ${\bf{X}} = {\bf{x}}^T \otimes {{\bf{I}}_M}$ with $	{{\bf{C}}^{i - 1}}$ defined by
\begin{equation}
%	\begin{small}
		\begin{aligned}
	{{\bf{C}}^{i - 1}} = \frac{1}{{{\sigma ^2}}}{\left( {{\bf E}^H\left( { {\bf{R}}_{{{\bf{X}}^{i - 1}}}^{ - 1} \odot \left({{\bf{1}}_{K \times K}} \otimes {{\bf{I}}_M}\right)} \right){\bf E}} \right)^T},
		\end{aligned}
%	\end{small}
\end{equation}
where ${\bf E} = {{{\bf{I}}_K} \otimes {{\bf{1}}_{M \times 1}}}$, $\bf A \otimes \bf B$ and $\bf A \odot \bf B$ represent the Kronecker product and the Hadamard product of $\bf A$ and $\bf B$, respectively, ${\bf I}_k$ denotes the identity matrix of size $k \times k$, and ${\bf{1}}_{{k_1} \times {k_2}}$ denotes the all-ones vector or matrix of size ${{k_1} \times {k_2}}$.   

On the other hand, concerning the term ${\bf{q}}_{\bf{X}}^H{\bf{R}}_{\bf{X}}^{ - 1}{{\bf{q}}_{\bf{X}}}$ which is jointly convex in ${{\bf{q}}_{\bf{X}}}$ and ${\bf{R}}_{\bf{X}}$, we obtain its lower bound via the first-order Taylor expansion as
\begin{equation} 
	\begin{small}
	\begin{aligned}		
		&{\bf{q}}_{\bf{X}}^H{\bf{R}}_{\bf{X}}^{ - 1}{{\bf{q}}_{\bf{X}}}\\
		&\geq  - {\rm{tr}}\left({\bf{R}}_{{{\bf{X}}^{i-1}}}^{ - 1}{{\bf{q}}_{{{\bf{X}}^{i-1}}}}{\bf{q}}_{{{\bf{X}}^{i-1}}}^H{\bf{R}}_{{{\bf{X}}^{i-1}}}^{ - 1}{{\bf{R}}_{\bf{X}}}\right)   \\ & \quad\quad\quad\quad +2{\mathop{\rm Re}\nolimits} \{ {\bf{q}}_{{{\bf{X}}^{i-1}}}^H{\bf{R}}_{{{\bf{X}}^{i-1}}}^{ - 1}{{\bf{q}}_{\bf{X}}}\} + {\bf{q}}_{{\bf{X}}^{i-1}}^H{\bf{R}}_{{\bf{X}}^{i-1}}^{ - 1}{{\bf{q}}_{{\bf{X}}^{i-1}}} \\
		%&= 2{\mathop{\rm Re}\nolimits} \{ tr\left( {{\bf{q}}_{{{\bf{X}}^i}}^H{\bf{R}}_{{{\bf{X}}^i}}^{ - 1}\left(\frac{1}{{{\sigma ^2}}}{{\bf{X}}^H}{\bf{y}} + {\bm{\Sigma }}_{\bf{h}}^{ - 1}{\bf{\hat h}})} )\}
		%\\&\quad- tr( {{\bf{R}}_{{{\bf{X}}^i}}^{ - 1}{{\bf{q}}_{{{\bf{X}}^i}}}{\bf{q}}_{{{\bf{X}}^i}}^H{\bf{R}}_{{{\bf{X}}^i}}^{ - 1}(\frac{1}{{{\sigma ^2}}}{{\bf{X}}^H}{\bf{X}} + {\bm{\Sigma }}_{\bf{h}}^{ - 1})} ) + {\rm{constant}}\\
		&\mathop  = \limits^{(a)} - \frac{1}{{{\sigma ^2}}}{\rm{tr}}\left({{\bf{B}}^{i-1}}{{\bf{X}}^H}{\bf{X}}\right)+\frac{2}{{{\sigma ^2}}}{\rm{tr}}\left({\rm Re}\{ {{\bf{A}}^{i-1}}{{\bf{X}}^H}\}\right)+ {\rm{constant}}\\
		&\mathop  = \limits^{(b)}  -{{\bf{x}}^H}{{\bf{F}}^{i-1}}{\bf{x}} + 2{\mathop{\rm Re}\nolimits} \{ {{\bf{x}}^H}{{\bf{d}}^{i-1}}\} + {\rm{constant}},
	\end{aligned}
\end{small}
\end{equation}
where (a) is obtained based on the definitions of ${{\bf{q}}_{\bf{X}}} = \frac{1}{{{\sigma ^2}}}{{\bf{X}}^H}{\bf{y}} + {\bm{\Sigma }}_{\bf{h}}^{ - 1}{\bf{\hat h}}$ and ${{\bf{R}}_{\bf{X}}} = \frac{1}{{{\sigma ^2}}}{{\bf{X}}^H}{\bf{X}} + {\bm{\Sigma }}_{\bf{h}}^{ - 1}$, and (b) is achieved by using ${\bf{X}} = {\bf{x}}^T \otimes {{\bf{I}}_M}$. In addition, ${{\bf{A}}^{i-1}} = {\bf{y}}{\bf{q}}_{{{\bf{X}}^{i-1}}}^H{\bf{R}}_{{{\bf{X}}^{i-1}}}^{ - 1}$,  ${{\bf{B}}^{{i-1}}} = {\bf{R}}_{{{\bf{X}}^{i-1}}}^{ - 1}{{\bf{q}}_{{{\bf{X}}^{i-1}}}}{\bf{q}}_{{{\bf{X}}^{i-1}}}^H{\bf{R}}_{{{\bf{X}}^{i-1}}}^{- 1}$, and ${{\bf{d}}^{i-1}}$ and ${{\bf{F}}^{i-1}}$ respectively take the forms:
	\begin{equation}
	{{\bf{d}}^{i - 1}} = \frac{1}{{{\sigma ^2}}}\left( {{{\bf{1}}_{1 \times M}}\left( {{{\bf{A}}^{i - 1}} \odot \left({{\bf{1}}_{1 \times K}} \otimes {{\bf{I}}_M}\right)} \right){\bf E}} \right)^T,
\end{equation}
and
	\begin{equation}
	{{\bf{F}}^{i - 1}} = \frac{1}{{{\sigma ^2}}}{\left( { {\bf E}^H\left( {{{\bf{B}}^{i - 1}}\ \odot \left({{\bf{1}}_{K \times K}} \otimes {{\bf{I}}_M}\right)} \right){\bf E}} \right)^T}.
\end{equation}

Combining (18) and (20), problem (16b) boils down to minimizing the surrogate upper bound function as follows:
\begin{equation} 
	\begin{aligned}
	{\bf{x}}^i= \mathop {\arg \min }\limits_{\bf{x}} 
		 {{\bf{x}}^H}{\left({{\bf{C}}^{i-1}}+{{\bf{F}}^{i-1}}\right)}{\bf{x}}
		 - 2{\mathop{\rm Re}\nolimits} \{ {{\bf{x}}^H}{{\bf{d}}^{i-1}}\} \\+
		\frac{\mu }{2}\left\| {{\bf{x}} + {\bm{\psi}}^{i-1}} \right\|_2^2,
	\end{aligned}
\end{equation}
where ${\bm{\psi}}^{i-1} = - \frac{1}{{{\alpha _Q}}}\sum\limits_{q = 1}^Q {{2^{q - 1}}{{\bf{v}}^{i}_q}}  + {\bm{\lambda }}^{i-1}$.
The objective function of problem (23) is convex quadratic w.r.t. $\bf x$ since ${{\bf{C}}^{i-1}}$ and ${{\bf{F}}^{i-1}}$ are both positive semidefinite and $\mu > 0$. Thus, the update of $\bf{x}$ can be easily obtained by
\begin{equation} 
{{\bf{x}}^i} = {\left(2\left({{{\bf{C}}^{i - 1}}}+{{{\bf{F}}^{i - 1}}}\right) + {\mu}{\bf{I}}_K\right)^{ - 1}}\left( { - \mu {{\bm{\psi}}^{i-1}} + 2{{\bf{d}}^{i - 1}}} \right).
\end{equation}

At this step, we are able to address problem (14) by iteratively calculating the closed-form expressions in (17), (24) and (16c) until the convergence is reached, which is also summarized in Algorithm \ref{RADMM}. Note that the convergence of ADMM to a general nonconvex problem remains an open issue, which is even harder with inexact optimization of the involved subproblems\cite{ADMM8933411}. In this work, we only obtain inexact solutions to subproblems (16a) and (16b). Nonetheless, our ADMM algorithm can still converge according to the empirical results. In particular, the above ADMM algorithm will be used for developing a neural network based detector via the deep unfolding technique in the next subsection.
	\begin{algorithm}[t]
%		\small
	\caption{Robust ADMM Detection Algorithm}
	\begin{algorithmic}[1]
		\REQUIRE $ {{\bf{y}}}$, ${\hat{\bf h}}$, and ${{\bm{\Sigma }}_{\bf{h}}}$.% and $I_{max}$.%The received signals $ {{\bf{y}}}$, the imperfect CSI ${\hat{\bf h}}$, the  covariance matrix of CSI uncertainty ${{\bm{\Sigma }}_{\bf{h}}}$ and the maximum number of iteration $I_{max}$.
		\STATE Initialize: ${\bf{x}}^{0}$, $\{ {{\bf{v}}_q^0}\} _{q = 1}^Q$, ${\bm{\lambda }}^{0}$, $i = 0$.
		\REPEAT 
		\STATE $i \gets i+1$.
		\STATE Sequentially update ${{\bf{v}}_q^i}$ for $q=1,2,\cdots,Q$ via (17).
		\STATE Update ${\bf{x}}^{i}$ via (24).
		\STATE Update ${\bm{\lambda }}^{i}$ via (16c).
		\UNTIL{convergence.}% \rmbf{or} $i = I_{max}$.}	
		\ENSURE  ${\hat{\bf x}}={\bf{x}}^{i}$.
	\end{algorithmic}
	\label{RADMM}
\end{algorithm}	

\subsection{Proposed RADMMNet}
In order to facilitate the proposed network design, we first perform some simplifications for the above ADMM algorithm. Specifically, the $MK$-dimensional matrix inversion ${\bf{R}}_{{{\bf{X}}^{i-1}}}^{ - 1}$ involved in the calculation of ${\bf C}^{i-1}$ in (24) requires a high complexity up to ${\cal O}\left({M^3}{K^3}\right)$, which can lead to a huge computational cost for both offline training and online calculation. To handle this, we further relax the upper bound in (18) by 
\begin{equation}
%	\begin{small}
	\begin{aligned}
		\ln \det \left({{\bf{R}}_{\bf{X}}}\right) &\leq
		{{\bf{x}}^H}{{\bf{C}}^{i-1}}{\bf{x}} + {\rm{constant}}\\
		&\leq {\varepsilon}^{i-1}{\left\|{\bf x}\right\|_2^2} + {\rm{constant}},
	\end{aligned}
%\end{small}
\end{equation}
where ${{\bf{C}}^{i - 1}}$ is given in (19) and $\varepsilon^{i-1}$ is set to satisfy ${\varepsilon^{i-1}}{{\bf I}_K} \succeq {{\bf{C}}^{i - 1}}$. We note that we regard ${\varepsilon^{i-1}}$ as a trainable parameter in the proposed network, whose value is determined after the network training is completed. In addition, when computing  ${\bf x}^i$ using (24), we need to calculate ${\bf{R}}_{{{\bf{X}}^{i-1}}}^{ - 1}{{\bf{q}}_{{{\bf{X}}^{i-1}}}}$ to obtain ${{\bf{d}}^{{i-1}}}$ and ${{\bf{F}}^{{i-1}}}$ (see (21) and (22)), which will also introduce a very high complexity of ${\cal O}\left({M^3}{K^3}+{M^2}{K^2}\right)$. { To avoid this, we adopt the conjugate gradient descent (CG) method to approximate ${\hat{\bf s}} = {\bf{R}}_{{{\bf{X}}^{i-1}}}^{ - 1}{{\bf{q}}_{{{\bf{X}}^{i-1}}}}$ by iteratively minimizing the quadratic function $f({\bf s}) = {{\bf{s}}^H}{\bf{R}}_{{{\bf{X}}^{i-1}}}{\bf{s}} - {{\bf{q}}_{{{\bf{X}}^{i-1}}}^H}{\bf{s}}$ \cite{CG123}.} 
\begin{algorithm}[t]
	%	\small
	\caption{CG Algorithm for Calculating ${\bf{R}}^{ - 1}{{\bf{q}}}$}
	\begin{algorithmic}[1]
		\REQUIRE ${\bf{R}}$ and ${{\bf{q}}}$.		
		\STATE Initialize: ${\bf{s}}_0$, ${{\bf{r}}_0} = {\bf q}-{\bf R}{{\bf{s}}_0}$, ${\bf{p}}_0 = {{\bf{r}}_0}$, $\xi = 0$.
		\REPEAT 
		\STATE $\xi \gets \xi+1$.
		\STATE ${\tau}_{\xi} = {{\bf r}_{\xi-1}^H{\bf r}_{\xi-1}}/{{\bf p}_{\xi-1}^H{\bf R}{\bf p}_{\xi-1}}$.
		\STATE ${\bf s}_{\xi} = {\bf s}_{\xi-1} + {\tau}_{\xi}{\bf p}_{\xi-1}$.
		\STATE ${\bf r}_{\xi} = {\bf r}_{\xi-1} - {\tau}_{\xi}{\bf R}{\bf p}_{\xi-1}$.
		\STATE ${\upsilon}_{\xi} = {{\bf r}_{\xi}^H{\bf r}_{\xi}}/{{\bf r}_{\xi-1}^H{\bf r}_{\xi-1}}$.
		\STATE ${\bf p}_{\xi} = {\bf r}_{\xi} + {\upsilon}_{\xi}{\bf p}_{\xi-1}$.
		\UNTIL{convergence.}
		\ENSURE ${\hat{\bf s}}={\bf{s}}_{\xi}$.
	\end{algorithmic}
	\label{CG}
\end{algorithm}	
{The detailed procedure of the CG method is described in Algorithm \ref{CG}, where the subscript ${{{\bf{X}}^{i-1}}}$ is temporarily omitted for the brevity of the notations.  Specifically, line 5, line 6, and line 8 represent the update of the solution ${\bf s}_{\xi}$, the residual ${\bf r}_{\xi}$, and the conjugate direction ${\bf p}_{\xi}$ in the $\xi$-th CG iteration,respectively, while line 4 and line 7 give the expressions of the corresponding step sizes ${\tau}_{\xi}$ and ${\upsilon}_{\xi}$. By resorting to the CG method, the complexity of computing ${\bf{R}}_{{{\bf{X}}^{i-1}}}^{ - 1}{{\bf{q}}_{{{\bf{X}}^{i-1}}}}$ is reduced to ${\cal O}\left(I_{\text{CG}}{M^2}{K^2}\right)$, where $I_{\text{CG}}$ denotes the number of CG iterations. Moreover, the iterative nature of the CG method is also friendly to the neural network design\footnote{{Although there exist other iterative methods for the same purpose, such as the Newton method and the steepest gradient descent method, they suffer from a high complexity and a slow convergence rate, respectively \cite{LCG}. As for the CG method, it only involves matrix multiplications per iteration with a much lower complexity. Meanwhile, the solution is updated along an independent conjugate direction with an exactly calculated step size in each CG iteration, which means that one will go to the extreme along one direction and there will be no repeated directions in the search procedure, leading to a faster convergence rate. Therefore, the CG method is more preferable.}}. On the other hand, it is noteworthy that a good initialization of ${\bf s}_0$ can improve the convergence of the CG method, which is generally set to $\bf 0$ if there is no {\emph{a prior}} knowledge. Fortunately, we find that the value of ${\bf{R}}_{{{\bf{X}}^{i}}}^{ - 1}{{\bf{q}}_{{{\bf{X}}^{i}}}}$ does not change dramatically as the iteration index $i$ increases. Therefore, when using the CG method to approximate ${\bf{R}}_{{{\bf{X}}^{i-1}}}^{ - 1}{{\bf{q}}_{{{\bf{X}}^{i-1}}}}$ in the $i$-th ADMM iteration ($i>1$), ${\bf s}_0$ can be initialized to ${\bf{R}}_{{{\bf{X}}^{i-2}}}^{ - 1}{{\bf{q}}_{{{\bf{X}}^{i-2}}}}$ obtained in the $(i-1)$-th ADMM iteration, which can efficiently reduce the required number of CG iterations down to $1$. This processing can also be regarded as transferring the intermediate variables to strength the connection between the network layers.}

	\begin{figure*}[t]
	{ \centering
		\includegraphics[width=13cm]{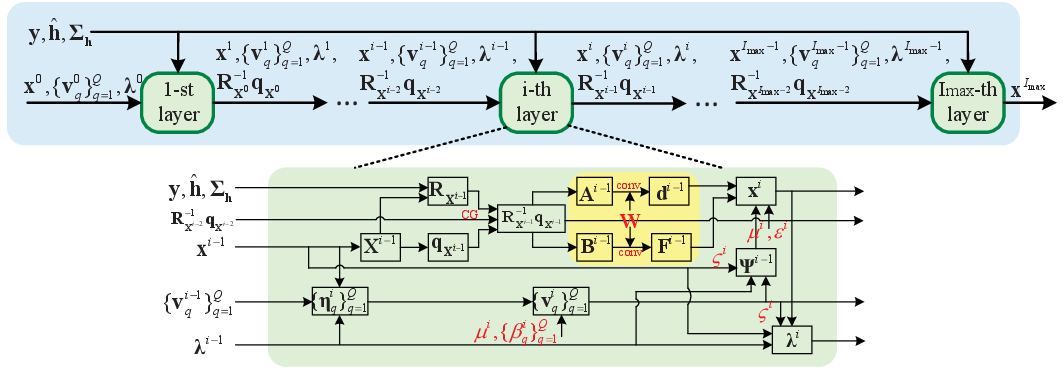}
		\caption{Block diagram of RADMMNet.}
		\label{fig2}}
\end{figure*}

After performing the above modifications for the proposed ADMM algorithm, we are ready to develop a neural network solution to the robust ML detection problem in (12). We employ the deep unfolding technique to represent the ADMM iterations by $I_\text{max}$ layers. Note that the involved penalty parameters $\mu$ and $\{ {{\beta}_q}\} _{q = 1}^Q$, as well as $\varepsilon$ involved in (25) can remarkably affect the performance and convergence of the proposed ADMM algorithm. In the proposed network, we set them to be trainable parameters which can be optimized via offline training, thus effectively avoiding a cumbersome numerical search. Furthermore, we adopt layer-wise parameters, i.e., $\{{\mu^i},{\{ {{\beta}^i_q}\} _{q = 1}^Q},{\varepsilon^i}\}_{i = 1}^{I_\text{max}}$, which can provide more degrees of freedom and potentially accelerate the convergence and improve the performance. 

{Note that (21) and (22) can be conveniently implemented in Tensorflow by convolutional layers with an identity matrix as the fixed filter. Instead of strictly following the derived expressions, we improve the learning ability of our network by replacing the fixed filter with a trainable one, so that the resulting deep unfolding network can also take advantage of the introduced data-driven structure. It yields:}
\begin{equation}
%	\begin{small}
		\begin{aligned}
	{{{\bf d}}^{i-1}} &= \frac{1}{{{\sigma ^2}}}\left( {{{\bf{1}}_{1 \times {M}}}\left( {{{{\bf{A}}}^{i - 1}} \odot \left({{\bf{1}}_{1 \times {K}}} \otimes {\bf W}\right)} \right){\bf E}} \right)^T \\&= \rm{conv}\left(\frac{1}{{{\sigma ^2}}}\left({{{\bf{A}}}^{i - 1}}\right)^T;{\bf W}\right),
		\end{aligned}
%\end{small}
\end{equation}
and
\begin{equation}
%		\begin{small}
		\begin{aligned}
	{{{\bf F}}^{i-1}} &= \frac{1}{{{\sigma ^2}}}{\left( {{\bf E}^H\left( { {{{\bf{B}}}^{i - 1}}\odot \left({{\bf{1}}_{{K} \times {K}}} \otimes {\bf W}\right)} \right){\bf E}} \right)^T} \\&= \rm{conv}\left(\frac{1}{{{\sigma ^2}}}\left({{{\bf{B}}}^{i - 1}}\right)^T;{\bf W}\right),
		\end{aligned}
%	\end{small}
\end{equation}
where ${\bf E} = {{{\bf{I}}_K} \otimes {{\bf{1}}_{M \times 1}}}$, and $\rm{conv\left(\cdot;{\bf W}\right)}$ means the convolutional operation with ${\bf W} \in {\mathbb R}^{M \times M}$ being the trainable filter. To guarantee the consistence of the output dimensions, the stride and the depth of the convolutional layers are set to $M$ and $1$, respectively. Unlike other layer-wise parameters, we share the same $\bf W$ for each network layer to limit the number of trainable parameters, which can effectively alleviate the training difficulty and avoid the overfitting problem.

To summarize, the $i$-th layer of the proposed network can be constructed as follows:
\begin{subequations}
	\label{d60}
	\begin{equation}
		{{{\bf{v}}}^i_q} = \mathop {\cal P}\nolimits_{{{\mathbb{B}}^K}} \left\{ \frac{{{2^{q - 1}}}{\mu^i}{{{\bm{\eta}}}_q^{i}}}{{{\mu^i} {4^{q - 1}} - 2\alpha _Q^2{\beta ^i_q}}} \right\},\quad q = 1,2, \cdots ,Q,
	\end{equation}
	\begin{equation}
		{{{\bf{x}}}^i} = {\left(2{{{\bf{F}}}^{i - 1}} + {\left({{\varepsilon}^i + {\mu}^i}\right)} {\bf{I}}_{K}\right)^{ - 1}}\left( { - {\mu^i} {{{\bm{\psi}}}^{i-1}} + 2{{{\bf{d}}}^{i - 1}}} \right),
	\end{equation}
	\begin{equation}
		{{{\bm{\lambda }}}^i} = {{{\bm{\lambda }}}^{i - 1}} + {{{\bf{x}}}^i} - \frac{1}{{{\alpha _Q}}}\sum\limits_{q = 1}^Q {{2^{q - 1}}{{\bf{v}}}_q^i},
	\end{equation}
\end{subequations}
where $ {{{\bm{\eta}}}_q^{i}} = { {\alpha _Q}\left({{{\bf{x}}}^{i-1}} + {{{\bm{\lambda }}}^{i - 1}}\right) -  \sum\limits_{p < q} {{2^{p - 1}}{{\bf{v}}}_p^{i}}  - \sum\limits_{p > q} {{2^{p - 1}}{{\bf{v}}}_p^{i-1}} }$, and ${{\bm{\psi}}}^{i-1} = -\frac{1}{{{\alpha _Q}}}\sum\limits_{q = 1}^Q {{2^{q - 1}}{{{\bf{v}}}^{i}_q}} + {{\bm{\lambda }}}^{i-1}$. 
Furthermore, in the expressions of ${{\bm{\psi}}}^{i-1}$ and (28c), we additionally introduce a trainable parameter $\varsigma^i$ by replacing the original term $- \frac{1}{{{\alpha _Q}}}\sum\limits_{q = 1}^Q {{2^{q - 1}}{{{\bf{v}}}^{i}_q}}$  with the term $-\frac{\varsigma^i}{{{\alpha _Q}}}\sum\limits_{q = 1}^Q {{2^{q - 1}}{{{\bf{v}}}^{i}_q}} - \left(1-\varsigma^i\right){{\bf x}^{i-1}}$. This corresponds to the over-relaxation scheme as suggested by \cite{ADMM8933411,ParasADMM}, which is expected to improve the convergence of ADMM iterations. 
Finally, the block diagram of the proposed RADMMNet is illustrated in Fig. \ref{fig2}.

\section{Low-Complexity Robust ADMM Detector Based Network Design}
Although we have bypassed the high-dimensional matrix inversion, RADMMNet can still be computationally intensive for the system with a large number of antennas, which imposes huge burdens on both offline and online processes. In this section, we derive a low-complexity robust ADMM detector and further provide the corresponding neural network design. 

\subsection{Low-Complexity Robust ADMM Detector}
To begin with, we rewrite (8) as
\begin{equation}
	{{\bf{y}}} = {\hat{\bf{H}}}{{\bf{x}}} + {{\bm \Delta}{\bf H}}{{\bf{x}}} + {{\bf{z}}},
\end{equation} 
where ${\hat{\bf H}}$ is the matrix form of ${\hat{\bf h}}$ and ${\bm \Delta}{\bf H} = {\bf H}-{\hat{\bf H}}$ is the corresponding zero-mean Gaussian distributed channel error. The row error covariance matrix of ${\bm \Delta}{\bf H}$, i.e., ${{\bm{\Sigma }}_{\bf{H}}} \triangleq {\mathbb{E}}\left\{{ {\left({\bm \Delta}{\bf H}\right)}{\left({\bm \Delta}{\bf H}\right)^H}} \right\}$, can be derived from ${{\bm{\Sigma }}_{\bf{h}}}$ in (7) by
\begin{equation}
	\begin{aligned} 
		{{\bm{\Sigma }}_{\bf{H}}} = \left({ {{\bf{1}}_{1 \times K}} \otimes {{\bf{I}}_{M}}}\right)\left( { {{{\bm{\Sigma }}_{\bf{h}}}}\odot \left({{\bf{I}}_{K}} \otimes {{\bf{1}}_{{M} \times {M}}}\right)} \right)\left({ {{\bf{1}}_{K \times 1}} \otimes {{\bf{I}}_{M}}}\right).
		%{{\bm{\Sigma }}_{\bf{H}}} = \sum\limits_{k = 1}^K { {{{\bm{\Sigma }}_{\bf{h}}}} \left(M(k - 1) + 1:Mk,M(k - 1) + 1:Mk\right)}.
	\end{aligned}
\end{equation} 

Based on (29), the key idea behind the low-complexity design is to impose a Gaussian approximation on the residual term ${\tilde{\bf r}} = {\bm \Delta}{\bf H}{{\bf{x}}} + {{\bf{z}}}$. That is, we assume that ${\tilde{\bf r}}$ follows a Gaussian distribution with zero mean and the covariance ${\bf{C}}_{\tilde{\bf r}} \triangleq {\mathbb{E}}\left\{{ {\left({\bm \Delta}{\bf H}{{\bf{x}}} + {{\bf{z}}}\right)}{\left({\bm \Delta}{\bf H}{{\bf{x}}} + {{\bf{z}}}\right)}^H}\right\} = {{\bm{\Sigma }}_{\bf{H}}} + {\sigma ^2}{{\bf{I}}_M}$, where the expectation is taken w.r.t. ${\bm \Delta}{\bf H}$, ${\bf z}$, and ${{\bf{x}}}$. Thus, the robust ML criterion can be simplified to
\begin{equation}
{\bf{\hat x}} = \mathop {\arg \min }\limits_{{\bf{x}} \in {{\cal{X}}^K}}  { {{\left({\bf{y}} - {\hat{\bf H}}{\bf x}\right)}^H}{{\bf{C}}_{\tilde{\bf r}}^{-1}}\left({\bf{y}} - {\hat{\bf H}}{\bf x}\right)}.
\end{equation}
%which can be interpreted as the ML criterion under colored Gaussian noises. 

We also resort to the ADMM framework to address problem (31), where the details are analogous to those in Section III-A. Specifically, following the reformulation of problem (14), we recast problem (31) by
\begin{equation} 
%	\begin{small}
	\begin{aligned}
		\mathop {\min }\limits_{{\bf{x}},\{ {{\bf{u}}_q},\forall q\} } &{{\left({\bf{y}} - {\hat{\bf H}}{\bf x}\right)}^H}{{\bf{C}}_{\tilde{\bf r}}^{-1}}\left({\bf{y}} - {\hat{\bf H}}{\bf x}\right) - \sum\limits_{q = 1}^Q {{\kappa _q}\left\| {{{\bf{u}}_q}} \right\|_2^2}
		\\{\rm{s.t.}}\quad &{\bf{x}} - \frac{1}{{{\alpha _Q}}}\sum\limits_{q = 1}^Q {{2^{q - 1}}{{\bf{u}}_q}}  = {\bf{0}},\\\quad & {\mathop{\rm Re}\nolimits} \{ {{\bf{u}}_q}\} ,{\mathop{\rm Im}\nolimits} \{ {{\bf{u}}_q}\}  \in {[- 1,1] ^K},q = 1, \cdots ,Q,\;
	\end{aligned}
%\end{small}
\end{equation}
where ${\kappa_q}>0,q = 1, \cdots ,Q $ are the penalty parameters. Then, the corresponding augmented Lagrangian function can be expressed as
\begin{equation}
%	\begin{small} 
	\begin{aligned}
		{L_\delta }\left({\bf{x}},\{ {{\bf{u}}_q}\}_{q = 1}^Q ,{\bm{\theta }}\right)= {{\left({\bf{y}} - {\hat{\bf H}}{\bf x}\right)}^H}{{\bf{C}}_{\tilde{\bf r}}^{-1}}\left({\bf{y}} - {\hat{\bf H}}{\bf x}\right) \\- \sum\limits_{q = 1}^Q {{\kappa _q}\left\| {{{\bf{u}}_q}} \right\|_2^2}  + \frac{\delta }{2}\left\| {{\bf{x}} - \frac{1}{{{\alpha _Q}}}\sum\limits_{q = 1}^Q {{2^{q - 1}}{{\bf{u}}_q}}  + {\bm{\theta}}} \right\|_2^2,
	\end{aligned}
%\end{small}
\end{equation}
where ${\bm{\theta}}$ and $\delta>0$ denote the scaled dual variables and the corresponding penalty parameter, respectively. Therefore, the $i$-th ADMM iteration can be given by
\begin{subequations}
	\begin{equation}
			\begin{aligned}
		\{ {{\bf{u}}_q^i}\} _{q = 1}^Q = \mathop {\arg \min }\limits_{{{\bf{u}}_q} \in {{\mathbb B}^K}} {L_\delta }\left({\bf{x}}^{i-1},\{ {{\bf{u}}_q}\}_{q = 1}^Q ,{\bm{\theta }}^{i-1}\right), %- \sum\limits_{q = 1}^Q {{\kappa _q}\left\| {{{\bf{u}}_q}} \right\|_2^2}  + \frac{\delta }{2}\left\| {{\bf{x}}^i - \frac{1}{{{\alpha _Q}}}\sum\limits_{q = 1}^Q {{2^{q - 1}}{{\bf{u}}_q}}  + {\bm{\theta}}^{i-1}} \right\|_2^2,
			\end{aligned}
	\end{equation}
	\begin{equation}
			\begin{aligned}
		{{\bf{x}}^i} = \mathop {\arg \min }\limits_{\bf{x}} {L_\delta }\left({\bf{x}},\{ {{\bf{u}}^{i}_q}\}_{q = 1}^Q ,{\bm{\theta }}^{i-1}\right), %{{\left({\bf{y}} - {\hat{\bf H}}{\bf x}\right)}^H}{{\bf{C}}_{\tilde{\bf r}}^{-1}}\left({\bf{y}} - {\hat{\bf H}}{\bf x}\right)+ \frac{\delta }{2}\left\| {{\bf{x}} - \frac{1}{{{\alpha _Q}}}\sum\limits_{q = 1}^Q {{2^{q - 1}}{{\bf{u}}_q^i}}  + {\bm{\theta}}^{i-1}} \right\|_2^2,
			\end{aligned}
	\end{equation}
%\vspace{-2mm}
	\begin{equation}
			\begin{aligned}
		{{\bm{\theta }}^i} = {{\bm{\theta }}^{i - 1}} + {{\bf{x}}^i} - \frac{1}{{{\alpha _Q}}}\sum\limits_{q = 1}^Q {{2^{q - 1}}{\bf{u}}_q^i}.
	\end{aligned}	
\end{equation}
\end{subequations}
Note that subproblem (34a) is almost the same as subproblem (16a) whose solution can be similarly achieved by
	\begin{equation}
	{\bf{u}}_q^{i} = \mathop {\cal P}\nolimits_{{{\mathbb{B}}^K}} \left\{\frac{{{2^{q - 1}}}{\delta}{{{{\bm{\omega}}}_q^{i}}}}{{\delta {4^{q - 1}} - 2\alpha _Q^2{\kappa_q}}} \right\}, \quad q = 1,2, \cdots ,Q,
\end{equation}
where $ {{{\bm{\omega}}}_q^{i}} = { {\alpha _Q}\left({{{\bf{x}}}^{i-1}} + {{{\bm{\theta }}}^{i - 1}}\right) -  \sum\limits_{p < q} {{2^{p - 1}}{{\bf{u}}}_p^{i}}  - \sum\limits_{p > q} {{2^{p - 1}}{{\bf{u}}}_p^{i-1}} }$. Moreover, by  dropping the terms independent of ${\bf x}$, we equivalently rewrite subproblem (34b) as  
\begin{equation} 
	\begin{aligned}
		{\bf{x}}^i= \mathop {\arg \min }\limits_{\bf{x}} 
		{{\bf{x}}^H}{\bm \Phi}{\bf{x}}
		- 2{\mathop{\rm Re}\nolimits} \{ {{\bf{x}}^H}{{\bm{\gamma}}^{i-1}}\},
	\end{aligned}
\end{equation}
where ${\bm \Phi} = {\hat{\bf H}}^H{{{\bf{C}}_{\tilde{\bf r}}^{-1}}}{\hat{\bf H}}+{\frac{\delta}{2}}{\bf I}_K$ and ${\bm{\gamma}}^{i-1} = {\hat{\bf H}}^H{{{\bf{C}}_{\tilde{\bf r}}^{-1}}}{\bf y} + \frac{\delta}{{2{\alpha _Q}}}\sum\limits_{q = 1}^Q {{2^{q - 1}}{{\bf{u}}^{i}_q}}  - \frac{\delta}{2}{\bm{\theta }}^{i-1}$. Since ${\bm \Phi}$ is positive definite, the optimal solution of $\bf{x}$ in each ADMM update is
\begin{equation} 
	{{\bf{x}}^i} = {{\bm \Phi}^{ - 1}}{\bm{\gamma}}^{i-1}.
\end{equation}

The above procedure is summarized in Algorithm \ref{LCRADMM}. Different from Algorithm \ref{RADMM} for problem (14), the update of $\bf{x}$ in Algorithm \ref{LCRADMM} is obtained in an exact closed form without the need to construct a surrogate function. The simplification can effectively reduce the computational complexity (see Section VI-C for details). 

\begin{algorithm}[t]
%	\small
		\arrayrulecolor{black}
		\color{black}
	\caption{Low-Complexity Robust ADMM Detection Algorithm}
	\begin{algorithmic}[1]
		\REQUIRE $ {{\bf{y}}}$, ${\hat{\bf H}}$, and ${{\bm{\Sigma }}_{\bf{H}}}$.% and $I_{max}$.%The received signals $ {{\bf{y}}}$, the imperfect CSI ${\hat{\bf h}}$, the  covariance matrix of CSI uncertainty ${{\bm{\Sigma }}_{\bf{e}}}$ and the maximum number of iteration $I_{max}$.
		\STATE Initialize: ${\bf{x}}^{0}$, $\{ {{\bf{u}}_q^0}\} _{q = 1}^Q$, ${\bm{\theta }}^{0}$, $i = 0$.
		\REPEAT 
		\STATE $i \gets i+1$.
		\STATE Sequentially update ${{\bf{u}}_q^i}$ for $q=1,2,\cdots,Q$ via (35).
		\STATE Update ${\bf{x}}^{i}$ via (37).
		\STATE Update ${\bm{\theta }}^{i}$ via (34c).
		\UNTIL{convergence.}% \rmbf{or} $i = I_{max}$.}	
		\ENSURE  ${\hat{\bf x}}={\bf{x}}^{i}$.
	\end{algorithmic}
	\label{LCRADMM}
\end{algorithm}	
	\begin{figure*}[t]
	\centering
	\includegraphics[width=13cm]{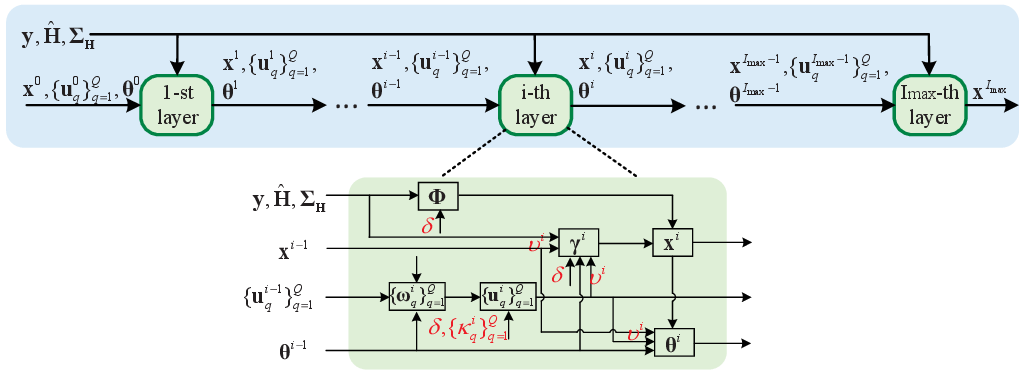}
	\caption{Block diagram of LCRADMMNet.}
	\label{fig3}
\end{figure*}
\subsection{Proposed LCRADMMNet}
Similar to RADMMNet, it is straightforward to build a model-driven network, referred to as LCRADMMNet, by unfolding the iterations of the derived low-complexity robust ADMM detection algorithm. {Note that ${{\bm \Phi}^{ - 1}} = \left({\hat{\bf H}}^H{{{\bf{C}}_{\tilde{\bf r}}^{-1}}}{\hat{\bf H}}+{\frac{\delta}{2}}{\bf I}_K\right)^{-1}$ in (37) involves an $M$-dimensional matrix inversion (${{{\bf{C}}_{\tilde{\bf r}}^{-1}}}$) and a $K$-dimensional matrix inversion (${{\bm \Phi}^{ - 1}}$), which, fortunately, only needs to be calculated  once during the ADMM iterations as long as $\delta$ remains constant. To inherit the complexity advantage, we set $\delta$ to be trainable and share it for different layers.} Besides, a set of layer-wise trainable parameters, denoted by $\{{\{ {{\kappa}^i_q}\} _{q = 1}^Q}\}_{i = 1}^{I_\text{max}}$, are used to take the role of the penalty parameters $\{ {{\kappa}_q}\} _{q = 1}^Q$ involved in Algorithm \ref{LCRADMM}. %Still, we have to convert the complex-valued variables into the equivalent real domain for the application of Tensorflow. Different from (26), the conversion should be based on the model in (35). We list the key real-valued representations in the following (${\bar{\bf y}}$ and ${\bar{\bf z}}$ are the same as in (26)):
%\begin{equation} 
%	\begin{small}
%		\begin{aligned}
%		{\bar{\bf y}} = \left[ {\begin{array}{*{20}{c}}
%				{{\mathop{\rm Re}\nolimits} \{ {\bf{y}}\} }\\
%				{{\mathop{\rm Im}\nolimits} \{ {\bf{y}}\} }
%		\end{array}} \right], 	{{\bar {{\bf x}}}} = \left[ {\begin{array}{*{20}{c}}
%					{{\mathop{\rm Re}\nolimits} \{ {\bf x}\} }\\
%					{{\mathop{\rm Im}\nolimits} \{ {\bf x}\} }
%			\end{array}} \right], {\bar{\hat{\bf H}}} = \left[ {\begin{array}{*{20}{c}}
%					{{\mathop{\rm Re}\nolimits} \{ {\hat{\bf H}}\} }&{ - {\mathop{\rm Im}\nolimits} \{ {\hat{\bf H}}\} }\\
%					{{\mathop{\rm Im}\nolimits} \{ {\hat{\bf H}}\} }&{{\mathop{\rm Re}\nolimits} \{ {\hat{\bf H}}\} }
%			\end{array}} \right], {{{\bf{\bar \Sigma }}}_{\bf{H}}} = \left[ {\begin{array}{*{20}{c}}
%					{{\mathop{\rm Re}\nolimits} \{ {{\bm{\Sigma }}_{\bf{H}}}\} }&{ - {\mathop{\rm Im}\nolimits} \{ {{\bm{\Sigma }}_{\bf{H}}}\} }\\
%					{{\mathop{\rm Im}\nolimits} \{ {{\bm{\Sigma }}_{\bf{H}}}\} }&{{\mathop{\rm Re}\nolimits} \{ {{\bm{\Sigma }}_{\bf{H}}}\} }
%			\end{array}} \right],	
%		\end{aligned}
%	\end{small}
%\end{equation}
%while other variables can be similarly acquired based on previous expressions. 

Therefore, the $i$-th layer of the proposed network can be expressed as:
\begin{subequations}
	\label{d60}
	\begin{equation}
%		\begin{small}
			\begin{aligned}
		{{{\bf{u}}}^i_q} = \mathop {\cal P}\nolimits_{{{\mathbb{B}}^K}} \left\{ \frac{{{2^{q - 1}}}{\delta}{{{\bm{\omega}}}_q^{i}}}{{{\delta} {4^{q - 1}} - 2\alpha _Q^2{\kappa ^i_q}}} \right\},\quad q = 1,2, \cdots ,Q,
			\end{aligned}
%	\end{small}
	\end{equation}
	\begin{equation}
%		\begin{small}
			\begin{aligned}
		{{{\bf{x}}}^i} = {{\left({{\bm \Phi}}\right)}^{ - 1}}{{\bm{\gamma}}}^{i-1},
			\end{aligned}
%	\end{small}
	\end{equation}
%\begin{equation}
%	{{{\bf{x}}}^i} = {\left({1-\upsilon^i}\right)}{{{\bf{x}}}^{i-1}} + {\upsilon^i}{{{\bf{x}}}^i},
%\end{equation}
	\begin{equation}
%		\begin{small}
			\begin{aligned}
			{{{\bm{\theta }}}^i} = {{{\bm{\theta }}}^{i - 1}} + {{{\bf{x}}}^i} - \frac{1}{{{\alpha _Q}}}\sum\limits_{q = 1}^Q {{2^{q - 1}}{{\bf{u}}}_q^i}, 
	\end{aligned}
%\end{small} 
	\end{equation}
\end{subequations}
where $ {{{\bm{\omega}}}_q^{i}} = { {\alpha _Q}\left({{{\bf{x}}}^{i-1}} + {{{\bm{\theta }}}^{i - 1}}\right) -  \sum\limits_{p < q} {{2^{p - 1}}{{\bf{u}}}_p^{i}}  - \sum\limits_{p > q} {{2^{p - 1}}{{\bf{u}}}_p^{i}} }$, ${{\bm \Phi}} = {{\hat{\bf H}}}^H{{{\bf{C}}_{\tilde{\bf r}}^{-1}}}{{\hat{\bf H}}}+{\frac{\delta}{2}}{\bf I}_K$, and ${{\bm{\gamma}}}^{i-1} = {{\hat{\bf H}}}^H{{{\bf{C}}_{\tilde{\bf r}}^{-1}}}{{\bf y}} + \frac{{\delta}}{{2{\alpha _Q}}}\sum\limits_{q = 1}^Q {{2^{q - 1}}{{{\bf{u}}}^{i}_q}} - \frac{\delta}{2}{{\bm{\theta }}}^{i-1}$. Here we can also apply the over-relaxation scheme by substituting $- \frac{1}{{{\alpha _Q}}}\sum\limits_{q = 1}^Q {{2^{q - 1}}{{\bf{u}}}_q^i}$ involved in the expressions of ${{\bm{\gamma}}}^{i-1} $ and (38c) with $- \frac{{\upsilon^i}}{{{\alpha _Q}}}\sum\limits_{q = 1}^Q {{2^{q - 1}}{{\bf{u}}}_q^i} - \left(1-\upsilon^i\right){{\bf x}^{i-1}}$,  where ${\upsilon^i}$ is the trainable relaxation parameter. We illustrate the proposed LCRADMMNet in Fig. \ref{fig3}.

\section{RDAKF-Based Channel Tracking}

We have developed two robust detection networks to suppress the performance deterioration caused by the imperfect CSI. However, the channel estimates of all data blocks are obtained only based on the pilots in the first data block. As such, the channel estimation errors can be remarkably accumulated for the subsequent data blocks without pilots, leading to degraded detection performance. To mitigate the error propagation, we now present a novel robust data-aided Kalman filter (RDAKF)-based channel tracking method, which incorporates not only the channel estimate acquired from the previous block as prior information but also the robust data estimates of the current block for further updating.

 The proposed channel tracking method can be described from the perspective of a Kalman filter, which consists of two key stages as detailed in the following.
%\begin{itemize}
%\item[1)] \textbf{One-step prediction stage}
\subsection{One-step prediction stage}
First, the state equation is utilized for one-step prediction. In this work, the channel variation model (1) can be regarded as the state equation. Then, the channel vector of the $n$-th block can be predicted as
\begin{equation} 
%		\begin{small}
	\begin{aligned}
		{{\hat{\bf h}}[n \left|{n - 1}\right.]} = {{\bm{\Lambda}}}{{\hat{\bf h}}}[n-1\left|{n - 1}\right.],\quad n = 2,3,\cdots,N+1,
	\end{aligned}
%\end{small}
\end{equation}
and the corresponding channel error covariance is given by
\begin{equation}
%		\begin{small}
	\begin{aligned}
		{\bm{\Sigma }}_{\bf{h}}[n \left|{n - 1}\right.] = {{\bm{\Lambda}}}{\bm{\Sigma }}_{\bf{h}}[n-1\left|{n - 1}\right.]{{\bm{\Lambda}}} + {\bar{\bm{\Lambda}}}{{\bf{C}}_{\bf h}}{\bar{\bm{\Lambda}}},\\ n = 2,3,\cdots,N+1,
	\end{aligned}
%\end{small}
\end{equation}
where ${\hat{\bf h}}[n \left|{n - 1}\right.]$ and ${\bm{\Sigma }}_{\bf{h}}[n \left|{n - 1}\right.]$ represent the \emph{a prior} channel estimate of the $n$-th block based on the previous estimate of the $(n-1)$-th block and the corresponding \emph{a prior} error covariance, respectively, while ${\hat{\bf h}}[{n-1} \left|{n - 1}\right.]$ and ${\bm{\Sigma }}_{\bf{h}}[n-1 \left|{n - 1}\right.]$ represent the \emph{a posterior} channel estimate of the ${(n-1)}$-th block obtained by utilizing the observation of the $(n-1)$-th block and the corresponding \emph{a posterior} error covariance, respectively. To guarantee the accuracy of the prediction, ${{\hat{\bf h}}[1]}$ and ${\bm{\Sigma }}_{\bf{h}}[1]$ acquired via (6) and (7) are used for the initialization of ${{\hat{\bf h}}[1\left|1\right.]}$ and ${\bm{\Sigma }}_{\bf{h}}[1\left|1\right.]$.

%\item[2)] \textbf{Updating stage}
\subsection{Updating stage}
We now refine the one-step prediction with the Kalman updating stage according to the observation equation in (10). Note that the measurement matrix in (10), i.e., ${\bf{X}}$, is unknown. To remedy this, we propose to reconstruct a measurement matrix using a robust MMSE detector \cite{Statis8944280}, which yields the data estimates as
\begin{equation}
%	\begin{small} 
	\begin{aligned} 
		{\hat{\bf x}}_t{[n\left| {n - 1} \right.]} = {{\hat{\bf H}}^H}{[n\left| {n - 1} \right.]}\Big({\hat{\bf H}}{[n\left| {n - 1} \right.]}{{{\hat{\bf H}}}^H}{[n\left| {n - 1} \right.]} \\+ {{\bm{\Sigma }}_{\bf{H}}}{[n\left| {n - 1} \right.]} + {\sigma ^2}{{\bf{I}}_M}\Big)^{ - 1}{\bf{y}}_t[n],
		%\\t = 1,2,\cdots,{L},n = 2,3,\cdots,N+1,
	\end{aligned}
%\end{small}
\end{equation}  
where ${\hat{\bf H}}[n \left|{n - 1}\right.]$ is the matrix form of ${\hat{\bf h}}[n \left|{n - 1}\right.]$ and ${{\bm{\Sigma }}_{\bf{H}}}[n \left|{n - 1}\right.]$ is the row covariance of the channel error ${\bm \Delta}{\bf H}[n \left|{n - 1}\right.] = {\bf H}[n \left|{n - 1}\right.]-{\hat{\bf H}[n \left|{n - 1}\right.]}$, which can be derived from ${{\bm{\Sigma }}_{\bf{h}}}[n \left|{n - 1}\right.]$ by (30). Based on (41), the row covariance of the data estimation error $ {\Delta{\bf x}}_t[n\left| {n - 1} \right.] = {\bf x}_t[n\left| {n - 1} \right.] - {\hat{\bf x}}_t[n\left| {n - 1} \right.]$ can be expressed by   
\begin{equation}
	\begin{small}
	\begin{aligned} 
		{{\bm{\Sigma }}_{{\bf{x}}_t}}[n\left| {n - 1} \right.] = {{\bf{I}}_K} - {{\bf{\hat H}}^H}[n\left| {n - 1} \right.]\Big({\bf{\hat H}}[n\left| {n - 1} \right.]{{{\bf{\hat H}}}^H}[n\left| {n - 1} \right.] \\ + {{\bm{\Sigma }}_{\bf{H}}}[n\left| {n - 1} \right.] + {\sigma ^2}{{\bf{I}}_M}\Big)^{ - 1}{\bf{\hat H}}[n\left| {n - 1} \right.],%\\		t = 1,2,\cdots,{L},n = 2,3,\cdots,N+1,
	\end{aligned}
\end{small}
\end{equation} 
where ${{\bm{\Sigma }}_{{\bf{x}}_t}}[n\left| {n - 1} \right.]$ is independent of the index $t$ and thus can be simply denoted by ${{\bm{\Sigma }}_{{\bf{x}}}}[n\left| {n - 1} \right.]$.

Using the estimate ${\hat{\bf x}}_t[n\left| {n - 1} \right.]$, the observation equation in (10) can be rewritten as
\begin{equation}
%\begin{small}
		\begin{aligned} 
		{{\bf{y}}_t[n]} = {\hat {\bf{X}}}_t[n\left| {n - 1} \right.]{{\bf{h}}[n]} + {{\tilde{\bf{z}}}_t[n]}, %\quad t = 1,2,\cdots,{L},n = 2,3,\cdots,N+1,
	\end{aligned}
%\end{small}
\end{equation} 
where the measurement matrix ${\hat{\bf{X}}}_t[n\left| {n - 1} \right.] = {\hat{\bf{x}}}_t^T[n\left| {n - 1} \right.] \otimes {{\bf{I}}_M}$, and the equivalent noise ${{\tilde{\bf{z}}}_t[n]} = \Delta{{\bf{X}}}_t^T[n\left| {n - 1} \right.]{{\bf{h}}[n]} + {{{\bf{z}}}_t[n]} = \left({\Delta{\bf{x}}}_t^T[n\left| {n - 1} \right.] \otimes {{\bf{I}}_M}\right){{\bf{h}}[n]} + {{{\bf{z}}}_t[n]}$ is assumed to be zero-mean Gaussian distributed with the covariance given by  
\begin{equation}
\begin{small}
	\begin{aligned} 
		{{\bf{C}}_{\tilde{\bf{z}}_t}}[n] &= {\mathbb{E}}\left\{ {\tilde{\bf z}}_t[n]{{{\tilde{\bf z}}}_t^H}[n]\right\}\\
		%&\mathop  = \limits^{(a)} {\mathbb{E}}\left\{ \Delta {{\bf{x}}^T}[n\left| {n - 1} \right.] \otimes {{\bf{I}}_M}{\bf{h}}[n]{{\bf{h}}^H}[n]\Delta {{\bf{x}}^*}[n\left| {n - 1} \right.] \otimes {{\bf{I}}_M}\right\}  + {\sigma ^2}{{\bf{I}}_M}\\
		&\mathop  = \limits^{(a)} {{{\mathbb{E}}\left\{ \Delta{{\bf{X}}}_t^T[n\left| {n - 1} \right.]{{\bf{C}}_{\bf{h}}}\left(\Delta{{\bf{X}}}_t^T[n\left| {n - 1} \right.]\right)^H\right\}}}  + {\sigma ^2}{{\bf{I}}_M}\\
		&\triangleq {{\bm{\Xi}}_t[n\left| {n - 1} \right.]} + {\sigma ^2}{{\bf{I}}_M},%\quad\quad t = 1,2,\cdots,{L},n = 2,3,\cdots,N+1,
	\end{aligned}
\end{small}
\end{equation}
\begin{figure*}[!t]
	\begin{equation}
		%	\begin{small}
		\begin{aligned}
			\left( {{{\bm{\Xi}}_t[n\left| {n - 1} \right.]}} \right)_{i,j} = {{\bf{1}}_{{1} \times {KM}}}{\left( {\left( {{{\bm{\Sigma }}_{\bf{x}}}[n\left| {n - 1} \right.] \otimes {\bf 0}_{{M} \times {M}}^{\backslash i,j}} \right) \odot {{\bf{C}}_{\bf{h}}}} \right)}{{\bf{1}}_{{KM} \times {1}}}, %\\	t = 1,2,\cdots,{L},n = 2,3,\cdots,N+1,
		\end{aligned}
		%\end{small}
	\end{equation}
	\begin{equation}
		%	\begin{small}
		\begin{aligned} 
			{{\bf{K}}^{(t)}}[n] = {{\bm{\Sigma }}_{\bf{h}}^{(t-1)}}[n\left| {n } \right.]{{\bf{\hat X}}_t}^H[n\left| {n - 1} \right.]\Big({{{\bf{C}}}_{\tilde{\bf{z}}}}[n]+ {{{\bf{\hat X}}}_t}[n\left| {n - 1} \right.]{{\bm{\Sigma }}_{\bf{h}}^{(t-1)}}[n\left| {n } \right.]{{{\bf{\hat X}}}_t}^H[n\left| {n - 1} \right.]\Big)^{ - 1},
		\end{aligned}
		%\end{small}
	\end{equation}
	\begin{equation}
		%		\begin{small}
		\begin{aligned} 
			{{\bf{\hat h}}^{(t)}}[n\left| n \right.] = {\bf{\hat h}}^{(t-1)}[n\left| {n } \right.] + {{\bf{K}}^{(t)}}[n]\Big({{\bf{y}}_t}[n] - {{{\bf{\hat X}}}_t}[n\left| {n - 1} \right.]{\bf{\hat h}}^{(t-1)}[n\left| {n } \right.]\Big),
		\end{aligned}
		%\end{small}
	\end{equation}
	\begin{equation}
		%		\begin{small}
		\begin{aligned} 
			{\bm{\Sigma }}_{{\bf{h}}}^{(t)}[n\left| n \right.] = \left( {{\bf{I}} - {{\bf{K}}^{(t)}}[n]{{{\bf{\hat X}}}_t}[n\left| {n-1 } \right.]} \right){{\bm{\Sigma }}_{\bf{h}}^{(t-1)}}[n\left| {n } \right.],
		\end{aligned} 
		%\end{small}
	\end{equation}
	\hrulefill
\end{figure*}where (a) is obtained by taking the expectation w.r.t. ${\bf h}[n]$ and ${{\bf{z}}}_t[n]$ simultaneously. Based on the definition of ${{\bm{\Sigma }}_{{\bf{x}}}}[n\left| {n - 1} \right.] = {\mathbb{E}}\left\{ \Delta {{\bf{x}}_t}[n\left| {n - 1} \right.]\Delta {{\bf{x}}_t^H}[n\left| {n - 1} \right.]\right\}$ and the property of Kronecker product, the ${(i,j)}$-th entry of ${{{\bm{\Xi}}_t[n\left| {n - 1} \right.]}}$ can be derived as (45) at the top of the next page, where ${\bf 0}_{{m_1} \times {m_2}}^{\backslash i,j}$ denotes the all-zeros matrix of size ${{m_1} \times {m_2}}$ except the $(i,j)$-th entry being $1$. Similar to ${{\bm{\Sigma }}_{{\bf{x}}}}[n\left| {n - 1} \right.]$, the index $t$ of ${{\bf{C}}_{\tilde{\bf{z}}_t}}[n]$ and  ${{{\bm{\Xi}}_t[n\left| {n - 1} \right.]}}$  can also be omitted since (44) and (45) are both independent of $t$.%and ${{\bf{C}}_{\bf{H}}}$ can be obtained from ${{\bf{C}}_{\bf{h}}}$ in the same way as (13).

\begin{algorithm}[t]
	%	\small
	\caption{RDAKF-Based Channel Tracking Algorithm}
	\begin{algorithmic}[1]
		\REQUIRE ${\bf{Y}}_P$, ${\bf{S}}_P$, ${\bm{\Lambda}}$, $\bar{\bm{\Lambda}}$, ${{\bf{C }}_{\bf{h}}}$, and ${{\bf{y}}_{t}[n]}$ for $t = 1,2, \cdots ,L$, $n = 2, \cdots ,N+1$.
		\ENSURE ${\hat{\bf x}}$.
		\STATE Initialize ${{\hat{\bf h}}[1\left|1\right.]}$ and ${\bm{\Sigma }}_{\bf{h}}[1\left|1\right.]$ via (6) and (7), respectively, $n = 1$.
		\FOR {$n = 2, \cdots ,N+1$}
		\STATE Predict ${{\hat{\bf h}}[n\left|n-1\right.]}$ and ${\bm{\Sigma }}_{\bf{h}}[n\left|n-1\right.]$ via (39) and (40), respectively.
		\STATE Calculate ${\hat{\bf x}}_t[n\left| {n - 1} \right.]$ for $t = 1,2, \cdots ,L$ via (41).
		\STATE Calculate  ${{\bf{C}}_{\tilde{\bf{z}}}}[n]$ via (44).
		\FOR {$t = 1,2, \cdots ,L$}
		\STATE Update ${{\bf{K}}^{(t)}}[n]$, ${{\hat{\bf h}}^{(t)}[n\left|n\right.]}$ and ${\bm{\Sigma }}_{\bf{h}}^{(t)}[n\left|n\right.]$ via (46)-(48), respectively.
		\ENDFOR
		\STATE ${{\hat{\bf h}}[n\left|n\right.]} \gets {{\hat{\bf h}}^{(L)}[n\left|n\right.]}$, ${\bm{\Sigma }}_{\bf{h}}[n\left|n\right.] \gets {\bm{\Sigma }}_{\bf{h}}^{(L)}[n\left|n\right.]$.
		\ENDFOR	
		\ENSURE $\left\{{{\hat{\bf h}}[n\left|n\right.]}\right\}_{n=1}^{N+1}$ and $\left\{{{\bm{\Sigma }}_{\bf{h}}[n\left|n\right.]}\right\}_{n=1}^{N+1}$.
	\end{algorithmic}
	\label{RDAKF}
\end{algorithm}

\begin{figure}[t]
	\centering
	\includegraphics[width=9cm]{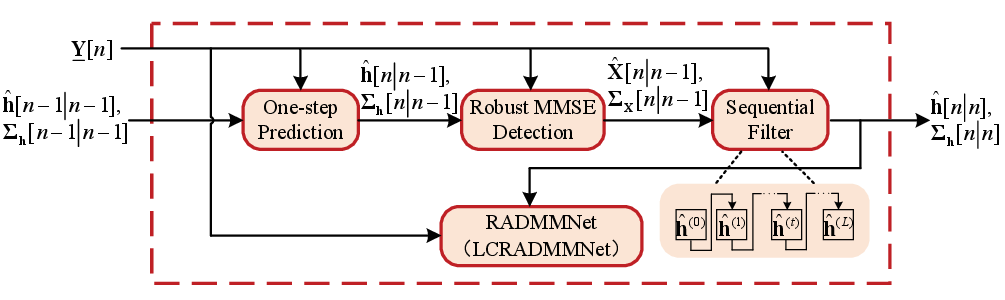}
	\caption{Block diagram of the proposed RDAKF-based receiver.}
	\label{fig4}
\end{figure}

Traditional Kalman updating can be readily performed by stacking the $L$ observation equations over a block into one $ML$-dimensional equation. However, the computational complexity to calculate the Kalman gain is up to ${\cal O}\left(M^3L^3\right)$ per data block. %\cite{BookEst}. 
 Note that the variables at different time slots can be considered uncorrelated with each other. Therefore we can update the estimates recursively for $t = 1,2,\cdots,L$ based on the sequential filter method \cite{9445013} via (46)-(48), where ${\bf{K}}$ is the Kalman gain matrix and ${(\cdot)}^{(t)}$ stands for the value of the input variable updated at time slot $t$. ${\bf{\hat h}}^{(0)}[n\left| {n } \right.]$ and ${{\bm{\Sigma }}_{\bf{h}}^{(0)}}[n\left| {n } \right.]$ are set to the one-step prediction ${\bf{\hat h}}[n\left| {n-1 } \right.]$ and ${{\bm{\Sigma }}_{\bf{h}}}[n\left| {n-1 } \right.]$, respectively. After completing the updating process along the $L$ time slots over a block, we obtain ${\bf{\hat h}}^{(L)}[n\left| {n } \right.]$ and ${{\bm{\Sigma }}_{\bf{h}}^{(L)}}[n\left| {n } \right.]$, which are then regarded as the \emph{a posterior} channel estimate and the corresponding error covariance, i.e, ${\bf{\hat h}}[n\left| {n } \right.]$ and ${{\bm{\Sigma }}_{\bf{h}}}[n\left| {n } \right.]$. It can be seen that the high-dimensional Kalman updating operation has been equivalently realized in a sequential manner, resulting in a lower complexity of ${\cal O}\left({M^3}L\right)$ per data block.
The overall procedure of the proposed channel tracking method is summarized in Algorithm \ref{RDAKF}. Based on this, we acquire the refined channel estimate and the corresponding error covariance, which can be readily utilized by RADMMNet and LCRADMMNet to improve the robust detection performance. The block diagram of the RDAKF-based receiver for the $n$-th block is shown in Fig. \ref{fig4}. 
%\end{itemize}

\section{Numerical Results}
%This section evaluates the performances of the proposed detection networks via numerical simulations. %The simulation settings and the training details are elaborated in the following. 
An uplink multiuser MIMO system is considered, where the number of the receive antennas at the base station and the number of the served users are set to $M =8$ and $K = 4$, respectively,  unless otherwise specified. The spatially correlated channels between the users and the base station are described by the Kronecker model \cite{Kron5338}, i.e., ${\bf{H}} = {\bf{R}}_r^{1/2}{{\bf{H}}_{{\rm{i}}{\rm{.i}}{\rm{.d}}}}{\bf{R}}_t^{1/2}$, where ${\bf H}_{\text{i.i.d}}$ is the i.i.d. Rayleigh fading channel matrix, each entry of which follows a Gaussian distribution with zero mean and unit variance, ${{\bf R}_t}$ and ${{\bf R}_r}$ are the exponential correlation matrices of the users and the receiver, which can be characterized by the correlation coefficients ${\rho_t}$ and ${\rho_r}$, respectively. In the simulation, we set ${\rho_t = 0.3}$, ${\rho_r = 0.7}$, and the length of a coherence block $L = 10$ with ${L_P} = K$. Without loss of generality, the channel temporal correlation coefficients of different users are assumed to be the same, i.e., ${\rho _1}={\rho _2}= \cdots ={\rho _K} = {\rho}$, where a lager $\rho$ means a slower channel variation. The signal-to-noise ratio (SNR) of the system is defined as ${\rm{SNR}} =\frac{K}{\sigma^2}$.

\begin{figure*}[!t]
	{ \subfigure[QPSK]
		{\begin{minipage}[t]{0.5\textwidth}
				\centering
				\includegraphics[width=7cm]{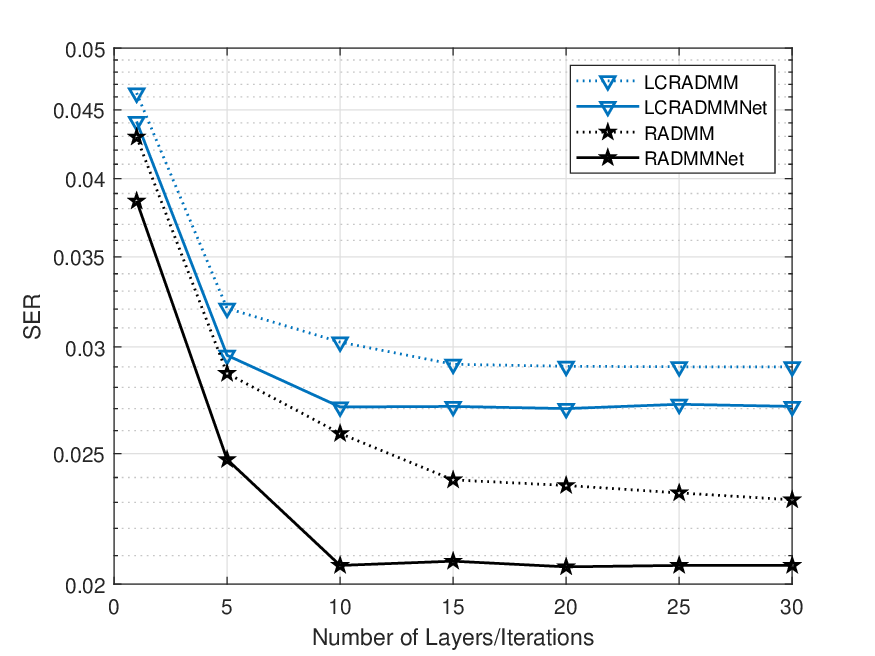}
			\end{minipage}		
		}
		\subfigure[16QAM]
		{\begin{minipage}[t]{0.5\textwidth}
				\centering
				\includegraphics[width=7cm]{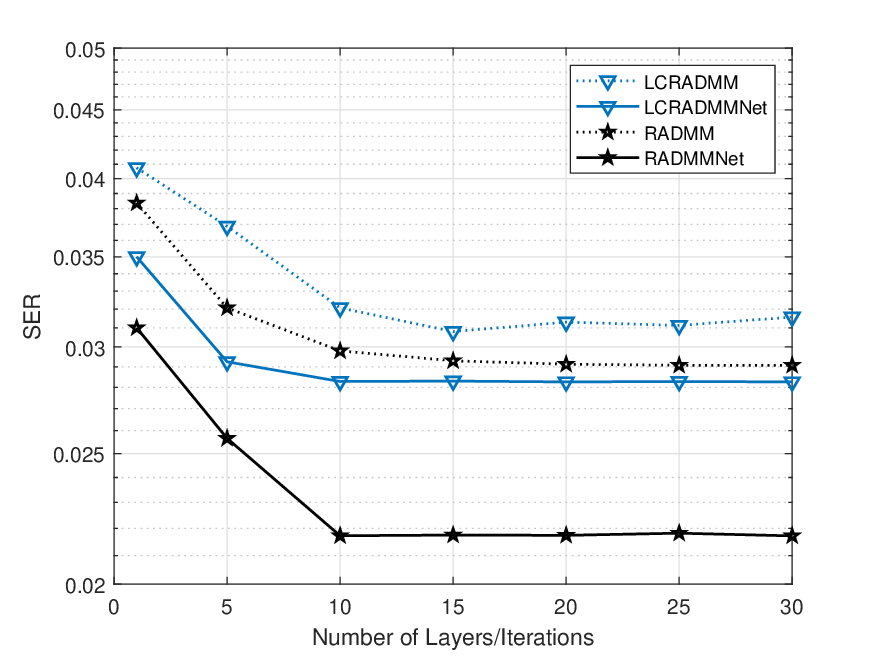}
			\end{minipage}		
		}	
		\caption{SER performances versus the number of layers under different modulation orders.}
		\label{fig5}}
\end{figure*}

%The details of the network training are provided as follows. Firstly, since the proposed RADMMNet has a complicated structure, performing a direct end-to-end training is time-consuming and can lead to a bad local minima. Hence, we propose to employ a two-step training strategy, which includes a layer-by-layer pre-training step and a fine-tuning step. We start with a one-layer RADMMNet and then successively add one layer to the tail of the last network. For each time, only the parameters of the current one layer are trained with the parameters of the previous layers being fixed. Finally, when we complete the training of all the $I_{\text{max}}$ layers, the parameters of the whole network will be further fine-tuned in an end-to-end manner. In fact, the layer-by-layer pre-training process can provide a good initial value for the parameters of each layer and thus facilitate the subsequent fine-tuning process. On the other hand, thanks to the small scale of the low-complexity LCRADMMNet, 
{Regarding the network training, both RADMMNet and LCRADMMNet are trained offline in an end-to-end manner with 10000 training samples and 2000 validation samples for each data block under each SNR point. We propose to employ a two-step training strategy, which includes a pre-training step and a fine-tuning step. The Adam optimizer is applied for the parameter learning with the initial learning rates of 0.01 for pre-training and 0.001 for fine-tuning. The batchsize for the two steps are set to 500 and 200, respectively, and the number of epochs are both set to 200. To prevent the overfitting and save the training time, we adopt the early-stopping mechanism, which will be triggered if the loss on the validation dataset does not decrease in successive 5 epochs. } In order to take the advantage of the supervised learning, we choose the MSE as the loss function for both networks, which is given by
\begin{equation}
%			\begin{small}
		\begin{aligned} 
	{\cal L}  = \frac{1}{{\left| {\cal S} \right|}}\sum\limits_{ ({{{\bf{x}}}^{{I_{\text{max} }}}},{{\bf{x}}}) \in \cal S} {\left\| {{{{\bf{x}}}^{{I_{\text{max} }}}} - {{\bf{x}}}} \right\|_2^2},
		\end{aligned}
%\end{small}
\end{equation} 
where ${{{\bf{x}}}^{{I_{\text{max}}}}}$ and ${{\bf{x}}}$ denote the network output and the true transmitted vector (also known as the label), respectively, and $\cal S$ is the training dataset. By minimizing the loss function, the deep unfolding networks are prompted to learn from the data in a supervised fashion, while maintaining the inherent mechanism of the original iterative algorithm. During the online simulation, we generate random testing samples continuously until 1000 symbol errors are collected so as to obtain more reliable SER results. On the other hand, the values of $L$, $N$, and other parameters used in the testing stage, are set to be identical as those in the training stage, which is reasonable since the system parameters such as the frame structure and the number of antennas are generally fixed in a given scenario. %Otherwise, the networks should be re-trained to adapt to a new scenario.

Apart from the proposed networks RADMMNet and LCRADMMNet, we also present the performances of the following baselines for comparisons: 1) the linear MMSE detector that directly uses imperfect CSI (Mismatched MMSE), 2) the ML detector that directly uses imperfect CSI in (9) (Mismatched ML), 3) The robust linear MMSE detector proposed in \cite{Statis8944280} (Robust MMSE), 4) The ADMM-based detection network proposed in \cite{Low1312} (ADMM-PSNet), 5) the inexact ADMM-based detection network proposed in \cite{LAN8715338} (ADMMNet), 6) the orthogonal approximate mesasge passing-based deep unfolding detection network proposed in \cite{OAMPNet} (OAMPNet), 7) the robust orthogonal approximate mesasge passing algorithm (ROAMP) and its corresponding deep unfolding detection network (ROAMPNet) \cite{Model8509622}\footnote{Note that ROAMP and ROAMPNet refer to OAMP and OAMPNet2 in \cite{Model8509622}, respectively. We rename them to emphasize that they belong to the robust designs and clearly distinguish them from the original algorithms OAMP and OAMPNet \cite{OAMPNet}, which are designed under the perfect CSI assumption.}, and 8) The robust ML detector in (12) (Robust ML).
%\begin{itemize}
%	\item The linear MMSE detector that directly uses imperfect CSI (Mismatched MMSE). 
%	\item The ML detector that directly uses imperfect CSI in (9) (Mismatched ML).
%	\item The robust linear MMSE detector proposed in \cite{Statis8944280} (Robust MMSE).
%	\item The ADMM-based detection network proposed in \cite{Low1312} (ADMM-PSNet).
%	\item { The inexact ADMM-based detection network proposed in \cite{LAN8715338} (ADMMNet).}
%	\item { The orthogonal approximate mesasge passing-based deep unfolding detection network proposed in \cite{OAMPNet} (OAMPNet).}
%	\item { The robust orthogonal approximate mesasge passing algorithm (ROAMP) and its corresponding deep unfolding detection network (ROAMPNet) \cite{Model8509622}\footnote{ Note that ROAMP and ROAMPNet refer to OAMP and OAMPNet2 in \cite{Model8509622}, respectively. We rename them to emphasize that they belong to the robust designs and clearly distinguish them from the original algorithms OAMP and OAMPNet \cite{OAMPNet}, which are designed under the perfect CSI assumption.}.}
%	\item The robust ML detector in (12) (Robust ML).
%\end{itemize}

\begin{remark} 
 Note that ADMM-PSNet, ADMMNet, and OAMPNet are all designed under the perfect CSI assumption but trained with imperfect CSI, while the proposed RADMMNet and LCRADMMNet are specially designed for the imperfect CSI case by explicitly incorporating the statistics of the imperfect CSI into the network structure. Consequently, both the proposed networks can exhibit stronger robustness in the presence of  CSI errors. As for the robust detection network ROAMPNet, it regards the sum of the CSI error and the receiver noise ${\tilde{\bf r}} = {\bm\Delta}{\bf H}{{\bf{x}}} + {{\bf{z}}}$ as an equivalent colored Gaussian noise, which is similar to the design principle of LCRADMMNet. However, LCRADMMNet requires a much lower complexity than ROAMPNet while with a comparable or even better performance. On the other hand, different from the above two robust detection networks that are developed based on a suboptimal robust ML metric using the Gaussian approximation, RADMMNet is established according to the optimal robust ML metric, which exploits not only the model-driven deep unfolding technique but also the introduced data-driven structure. Therefore, it can achieve better performance than LCRADMMNet and ROAMPNet. The advantages of the proposed networks over the existing DL structures will be validated by the following numerical results.
 \end{remark}

\begin{figure*}[t]
	\centering
	\includegraphics[width=\linewidth]{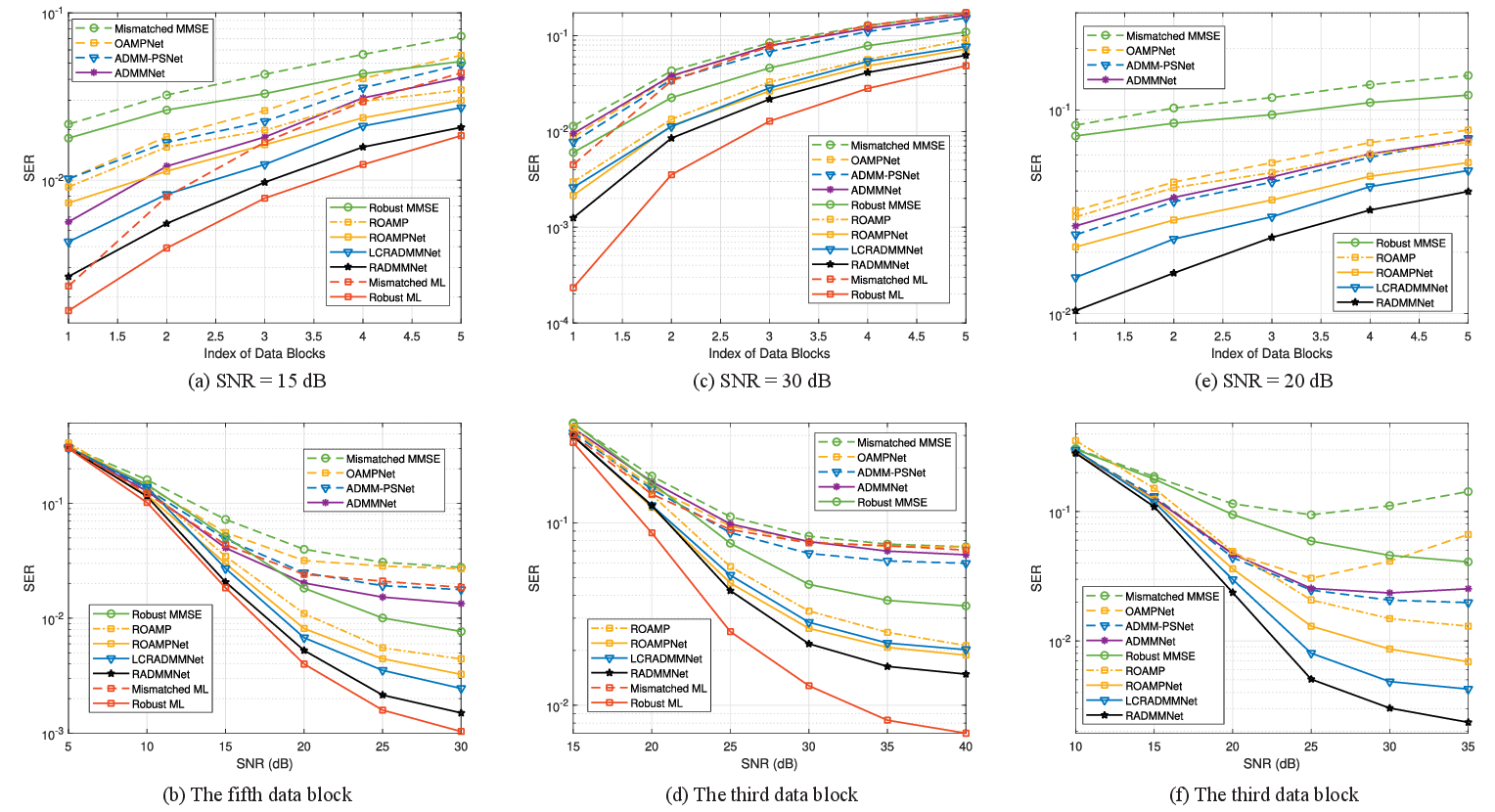}
	\caption{SER performances with LMMSE channel estimation for the scenario of (a)(b) $M=8$, $K=4$, and $\rho = 0.99$ under QPSK modulation, (c)(d) $M=8$, $K=4$, and $\rho = 0.995$ under 16QAM modulation, and (e)(f) $M=8$, $K=8$, and $\rho = 0.995$ under QPSK modulation.}
	\label{fig6}
\end{figure*}

	\subsection{Convergence Property}
Different from iterative algorithms where the number of iterations can be adjusted adaptively, the numbers of layers for deep unfolding-based networks needs to be pre-determined. For this purpose, we first investigate the convergence property of the proposed RADMMNet and LCRADMMNet. Fig. \ref{fig5} plots the symbol error rate (SER) performances of the proposed networks versus the number of layers under different modulation orders, where the corresponding two ADMM-based algorithms, denoted by ``RADMM'' and ``LCRADMM'', are also provided for comparison. {Note that in the model considered in our work, the channel variation becomes more serious as the data index increases or $\rho$ gets smaller. Meanwhile, it is intuitive that higher-order modulated signals are more vulnerable to the channel variation, which means that a milder condition is required in the 16QAM case to guarantee an acceptable performance. Therefore, for the QPSK case, we take the fifth data block with SNR = 15 dB and $\rho = 0.99$ for illustration, while for the 16-QAM case, we consider the third data block with SNR = 30 dB and $\rho = 0.995$. It can be observed that both RADMMNet and LCRADMMNet can considerably outperform the corrresponding ADMM-based algorithms with faster convergence rates, i.e., within 10 layers, indicating the effectiveness of the designed network and the advantage of the parameters learned via offline training.  Meanwhile, we would like to note that, for the same number of layers/iterations, LCRADMMNet requires nearly the same level of complexity as LCRADMM, while RADMMNet can efficiently reduce the complexity of RADMM due to the introduced simplifications. Consequently, in the following simulation, we set the number of layers of both RADMMNet and LCRADMMNet to 10 and do not present the results of RADMM and LCRADMM for the simplicity of the curves.} We also mention that the number of CG iterations involved in the first layer of RADMMNet, i.e., $I_{\text{CG}}$, is fixed to 15 empirically, which can reach almost the same SER performance as that with an exact matrix inversion.

	\subsection{Performance Evaluation}
In this subsection, we compare the performances of different detectors using the LMMSE channel estimation presented in Section II and the RDAKF-based channel tracking presented in Section IV, respectively. Then, {the performances with larger MIMO systems are also provided.}%, and the realistic channels are also used to test the proposed networks.}
\subsubsection{LMMSE channel estimation}
We first consider the case that the imperfect CSI is acquired by the LMMSE channel estimation. Fig. 6(a) shows the SER performances of different detection algorithms versus the index of data blocks for SNR = 15 dB and $\rho = 0.99$ under QPSK modulation. Since the channel estimates of all data blocks are obtained using the pilot in the first block, the channel errors will be accumulated due to the time variation of channels. As a result, the performances of all the detection algorithms degrade as the index increases, among which even the mismatched ML detector suffers from an unbearable performance loss. In contrast, the robust detectors can effectively compensate the deterioration caused by the channel aging and achieve satisfying performance, validating the necessity of the robust design. In particular, the proposed RADMMNet yields a relatively low SER and approaches the optimal robust ML detector. LCRADMMNet performs a little worse than RADMMNet while with lower complexity (see Section VII-C). With the significant performance advantage of RADMMNet and LCRADMMNet over the non-robust detectors, more data blocks can be transmitted with only one pilot block, which can improve the spectrum efficiency.

\begin{figure*}[!t]
	\centering
	\includegraphics[width=\linewidth]{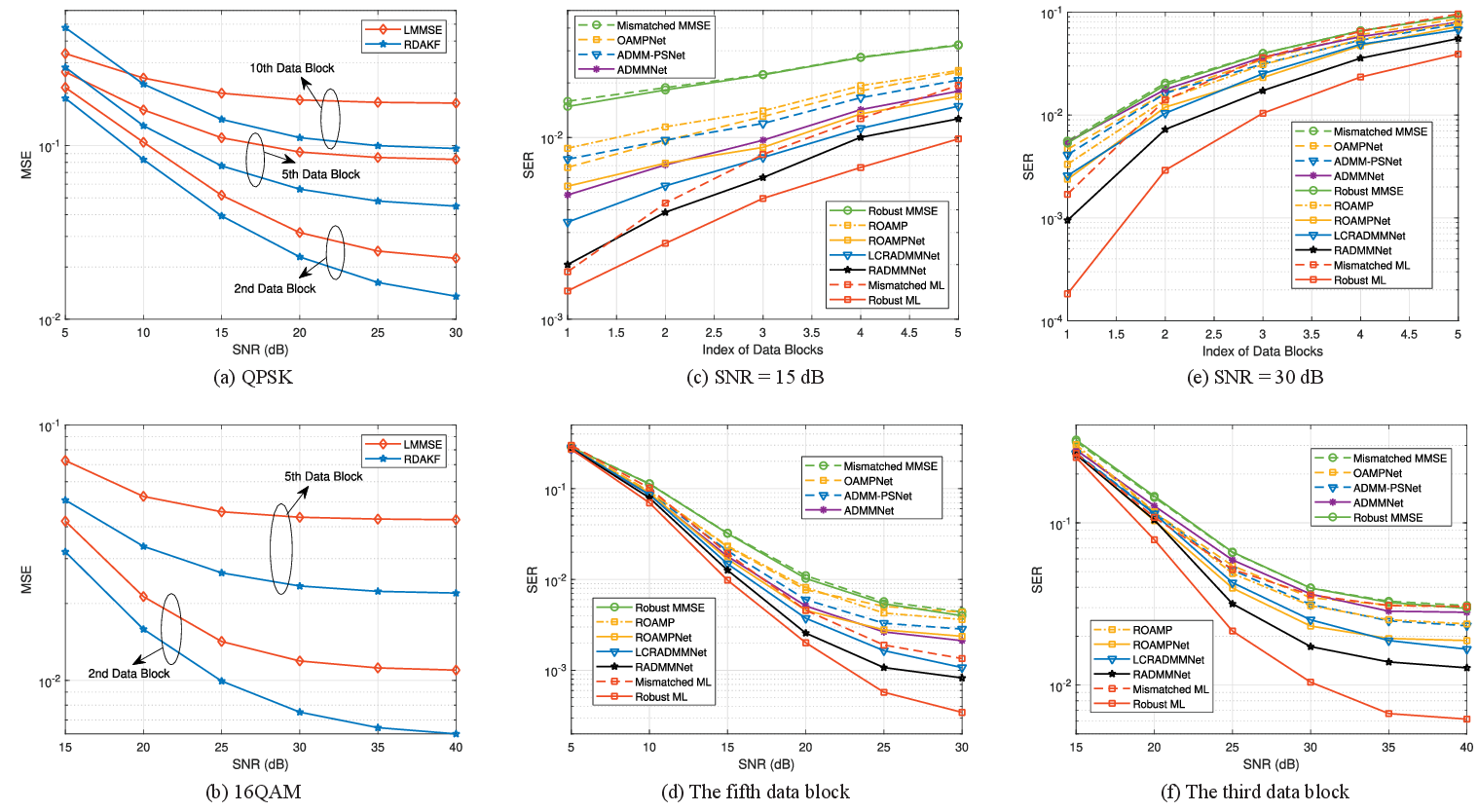}
	\caption{(a)(b) NMSE performances for different indices of data blocks for the scenario of $M=8$ and $K=4$ under different modulation orders, and SER performances with RDAKF channel tracking for the scenario of (c)(d) $M=8$, $K=4$, and $\rho = 0.99$ under QPSK modulation and (e)(f) $M=8$, $K=4$, and $\rho = 0.995$ under 16QAM modulation.}
	\label{fig7}
\end{figure*}

The SER performances versus SNR for the fifth data block are illustrated in Fig. 6(b). As can be seen, the detectors that neglect the imperfection of CSI suffer from high error floors in the high-SNR region. {We notice that both ADMM-PSNet and ADMMNet show slightly better performances than the mismatched ML detector. The reason is that they are trained to approach the true transmitted vectors using the imperfect CSI via supervised learning. In this way, the embedded statistical information of CSI errors can be implicitly utilized, which consequently provides the networks partial robustness against imperfect CSI.}  On the other hand, by fully incorporating the statistical information of the imperfect CSI into the network design, the proposed RADMMNet exhibits a much better performance. In addition, the lower-complexity LCRADMMNet can also considerably outperform other baselines including the existed non-robust detectors and the robust ones.

Next we present the SER performances under 16-QAM constellation with $\rho = 0.995$, which corresponds to a lower-mobility scenario. Figs. 6(c) and 6(d) present the SER performances versus the index of data blocks for SNR = 30 dB and those versus SNR for the third data block, respectively. It can be observed that, even with such a slower channel variation, the increasing index will lead to a serious performance degradation for each detection algorithm. Hence, in the 16-QAM case, the pilot block can only be followed by a few data blocks in a frame to guarantee the performance. Nevertheless, the proposed robust detection networks still perform much better than the non-robust detectors. {Besides, as shown in Fig. 6(d), compared with the robust MMSE detector, noticable performance gains up to 12 dB and 15 dB in the high-SNR region can be achieved by RADMMNet and LCRADMMNet, respectively. Although the performances of LCRADMMNet and ROAMPNet are close, the complexity of LCRADMMNet is much lower than that of ROAMPNet, which will be analyzed in detail in Section~VII-C.}

To verify the advantages of our proposed networks in the challenging scenario when the ratio of the number of receive antennas to that of transmit antennas equals 1, we conduct simulations by setting $M = K = 8$ and $\rho = 0.995$ under the QPSK modulation. The SER performances versus the index of data blocks for SNR = 20 dB and those versus SNR for the third data block are plotted in Figs. 6(e) and 6(f), respectively, where the mismatched ML detector and the robust ML detector are not incorporated due to their high complexities. Compared with the previous results with $M = 8$ and $K = 4$ in Figs. 6(a) and 6(b), the performance gains of the proposed RADMMNet and LCRADMMNet over other baselines, especially the mismatched MMSE detector and the robust MMSE detector, are more distinct. We also note that, for the high SNR region where the CSI error is the dominant factor of deteriorating the performance, the non-robust detectors and detection networks suffer from an error floor or even an abnormally rising SER, while the proposed robust detection networks can still yield satisfying performances.

\begin{figure*}[!t]
	{\subfigure[$M = 32$, $K = 8$, and $\rho = 0.98$]
		{\begin{minipage}[t]{0.5\textwidth}
				\centering
				\includegraphics[width=7cm]{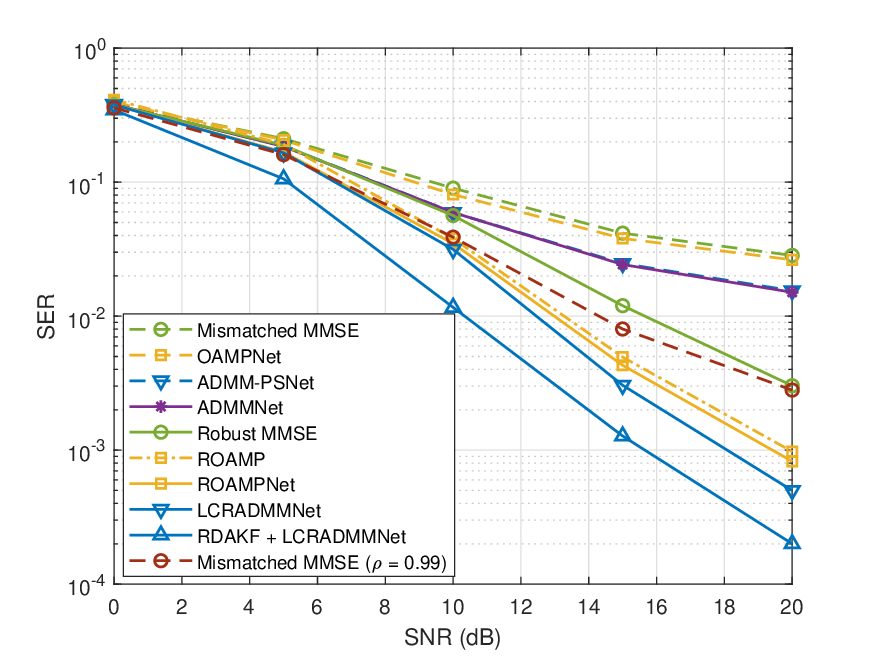}
			\end{minipage}		
		}
		\subfigure[$M = 128$, $K = 16$, and $\rho = 0.95$]
		{\begin{minipage}[t]{0.5\textwidth}
				\centering
				\includegraphics[width=7cm]{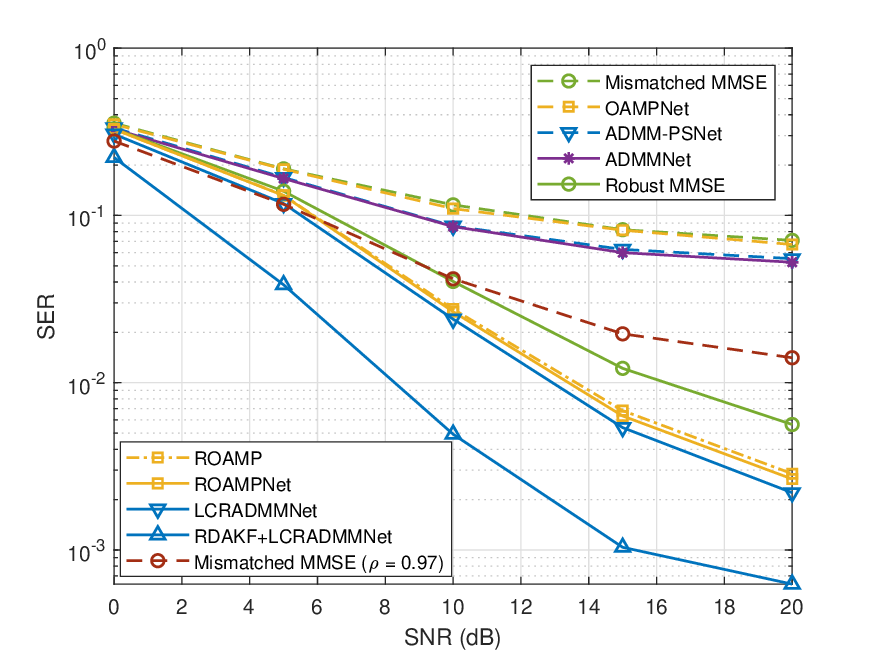}
			\end{minipage}		
		}	
		\caption{SER performances versus SNR for the fifth data block under QPSK modulation in large-scale MIMO scenarios.}
		\label{fig12}}
\end{figure*}

\subsubsection{RDAKF-based channel tracking}
As an improved alternative of the LMMSE channel estimation, the RDAKF-based channel tracking method is evaluated in terms of the NMSE performances of channel estimates and the SER performances, respectively, where the NMSE is defined as $\frac{{\left\| {{\hat{\bf{H}}} - {\bf{H}}} \right\|_F^2}}{{\left\| {\bf{H}} \right\|_F^2}}$.
Figs. 7(a) and 7(b) compare the NMSE performances between the two CSI acquisition methods for different indices of data blocks under different modulation orders. As expected, the RDAKF-based method can effectively track the channel variation with the aid of the pre-estimated data symbols in both the QSPK and 16-QAM cases. For the larger index of data blocks, the estimation performance gap between the RDAKF and the LMMSE-based methods becomes more evident. This is due to the fact that the error propagation caused by the coarse prediction of the LMMSE method gets more severe, which can be suppressed in the updating stage of the RDAKF method. We also notice from Fig. 7(a) that, in the low-SNR region, the inaccurate data estimates may deteriorate the accuracy of channel estimates, leading to an even poorer NMSE performance of the RDAKF method than that of the LMMSE-based method. 

The SER performances versus the index of data blocks are plotted in Fig. 7(c) with the same simulation settings as in Fig. 6(a). Based on the refined CSI, all the detection algorithms enjoy a performance improvement compared with the results in Fig. 6(a). Therefore, a frame structure containing more data blocks can be supported with the RDAKF-based channel tracking method. However, there is almost no advantage of the robust MMSE-based detector over the mismatched MMSE-based one, indicating the limitation of a linear robust design when the channel estimate is relatively accurate. Besides, the performance gaps between non-robust detection networks and LCRADMMNet are narrowed with the reduced channel error. Fig. 7(d) shows the performances versus SNR for the fifth data block. High error floors can be observed for the non-robust detetors as well as the robust MMSE detector, since they are unable to make good use of the relatively small CSI uncertainty. In contrast, RADMMNet still shows a comparable performance to the robust ML detector by fully exploiting the statistics of CSI errors. 

Similar trends can be found in the 16-QAM case as shown in Figs. 7(e) and 7(f), where the simulation settings are the same as in Figs. 6(c) and 6(d), respectively. The proposed RADMMNet can maintain the superiority over other baselines, demonstrating its stronger robustness against imperfect CSI. However, the SER performance gains obtained by utilizing the RDAKF-based channel tracking method are less obvious than those in the QPSK case, albeit with the similar level of improvement in the NMSE performance as illustrated in Fig. 7(b). This is because the detection of 16-QAM signals requires more accurate channel estimates. In addition, the performance gap from the robust ML detector is more evident than in the QPSK case, since 16-QAM signals are more susceptible to CSI errors. Hence, we conclude that the proposed RDAKF-based method is more applicable for QPSK modulation. {The results for the more challenging scenario of $M = K = 8$, corresponding to the simulation settings in Figs. 6(e) and 6(f), also exhibit a similar behavior with improved performances benefited from the RDAKF method, which are not presented here due to the space limitation.}

\begin{table*}[!t]
	\caption{Complexity Comparison of Different Detectors}
	%	\arrayrulecolor{blue}
%		\renewcommand\arraystretch{1}
	\resizebox{\textwidth}{!}{%
		\begin{tabular}{|c|c|ccc|}
			\hline
			\multirow{2}{*}{Detectors} & \multirow{2}{*}{Complexity} & \multicolumn{3}{c|}{Average CPU time (s)}                                                 \\ \cline{3-5} 
			&                             & \multicolumn{1}{c|}{$M=8$, $K=4$, QPSK} & \multicolumn{1}{c|}{$M=8$, $K=4$, 16QAM} & $M=128$, $K=16$, QPSK \\ \hline
			Mismached MMSE                       &        ${\cal{O}}\left({K^3}\right)$                     &   \multicolumn{1}{c|}{2.60e-5}             & \multicolumn{1}{c|}{2.68e-5}              &    1.14e-4             \\ \hline
			
			Robust MMSE                      &    ${\cal{O}}\left({M^3}\right)$                         & \multicolumn{1}{c|}{3.62e-5}             & \multicolumn{1}{c|}{3.71e-5}              &        2.27e-3         \\ \hline
			
			OAMPNet                    &        ${\cal{O}}\left({I_{\rm {max}}}{K^3}\right)$                     & \multicolumn{1}{c|}{5.41e-4}             & \multicolumn{1}{c|}{6.30e-4}              &       3.42e-3          \\ \hline
			
			ADMM-PSNet                 &           ${\cal{O}}\left({K^3}+{I_{\rm {max}}}{K^2}\right)$                  & \multicolumn{1}{c|}{1.15e-4}             & \multicolumn{1}{c|}{1.62e-4}              &     4.52e-4            \\ \hline
			
			ADMMNet                    &         ${\cal{O}}\left({I_{\rm {max}}}\left(MK + K^2\right)\right)$                    & \multicolumn{1}{c|}{1.21e-4}             & \multicolumn{1}{c|}{1.75e-4}              &     5.38e-4            \\ \hline
			
			ROAMPNet                   &             ${\cal{O}}\left({I_{\rm {max}}}{M^3}\right)$                & \multicolumn{1}{c|}{6.85e-4}             & \multicolumn{1}{c|}{7.75e-4}              &      1.29e-2           \\ \hline
			
			LCRADMMNet                &         ${\cal{O}}\left({M^3}+{K^3}+{I_{\rm {max}}}{K^2}\right)$                    & \multicolumn{1}{c|}{{1.32e-4}}             & \multicolumn{1}{c|}{1.98e-4}              &      {2.72e-3}           \\ \hline
			
			RADMMNet                   &   ${\cal{O}}\left({\left(I_{\rm{CG}}+I_{\rm {max}}-1\right){{M^2}{K^2}}}+I_{\rm {max}}{K^3}\right)$                         & \multicolumn{1}{c|}{1.48e-3}             & \multicolumn{1}{c|}{1.73e-3}              &       {/}           \\ \hline
			
			Mismatched ML              &        ${\cal{O}}\left(4^{{QK}}{K^3}\right)$                     & \multicolumn{1}{c|}{4.52e-4}             & \multicolumn{1}{c|}{7.20e-2}              &           /       \\ \hline
			
			Robust ML                  &         ${\cal{O}}\left(4^{{QK}}{{M^3}{K^3}}\right)$                    & \multicolumn{1}{c|}{3.41e-2}             & \multicolumn{1}{c|}{8.43}              &       /        \\ \hline
		\end{tabular}%
	}
\end{table*}

\subsubsection{Large-scale MIMO} {We now investigate the SER performances for MIMO system with larger sizes. Since a larger array can effectively make up the performance loss caused by the imperfect CSI, it can be inferred that a faster channel variation can be well handled with more receive antennas. To validate this, we test the detection algorithms under higher-speed scenarios by setting smaller $\rho$'s for larger-scale MIMO systems, i.e., $\rho = 0.98$ for  $M = 32$ and $K = 8$, and $\rho = 0.95$ for  $M = 128$ and $K = 16$ in Figs. \ref{fig12}(a) and \ref{fig12}(b), respectively. The results of the mismatched ML detector, the robust ML detector, and RADMMNet are not presented due to their unaffordable complexities. It can be observed that the proposed LCRADMMNet clearly outperforms most benchmark schemes and even the MMSE detector with larger $\rho$'s. Additionally, LCRADMMNet can yield a slightly better performance  than ROAMPNet while with a much lower complexity, which will be illustrated in the next subsection. On the other hand, the performance gains brought by the RDAKF-based channel tracking are conspicuous, especially for the scenario of $M = 128$ and $K = 16$, thanks to the large antenna array gain that results in a high probability of obtaining correct pre-esitmated symbols.}

\subsection{{Complexity Analysis}}
{The computational complexities of different detectors are analyzed in Table II. For the proposed RADMMNet, the major computations lie in the CG iterations and the matrix inversion{\footnote{Note that the first layer of RADMMNet requires a complete CG procedure with $I_{\rm{CG}}$ CG iterations, while the subsequent $I_{\rm{max}}-1$ layers only involve one CG iteration per layer, resulting in $I_{\rm{CG}}+I_{\rm{max}}-1$ CG iterations in total.}}, which leads to a complexity of  ${\cal{O}}\left({\left(I_{\rm{CG}}+I_{\rm{max}}-1\right){{M^2}{K^2}}}+I_{\rm{max}}{K^3}\right)$, while the other proposed network LCRADMMNet only needs to calculate one $M$-dimensional matrix inversion, one $K$-dimensional matrix inversion, and a few matrix multiplications with a total  complexity of ${\cal{O}}\left(M^3 + K^3 + {I_{\rm{max}}}K^2\right)$. The robust detection network ROAMPNet, as one of the baselines, requires a higher complexity of ${\cal{O}}\left({I_{\rm {max}}}{M^3}\right)$. On the other hand, the complexities of the mismatched ML detector and the robust ML detector  are ${\cal{O}}\left(4^{{QK}}{K^3}\right)$ and ${\cal{O}}\left(4^{{QK}}{{M^3}{K^3}}\right)$, respectively, which can be prohibitively high for a high-modulation order or a large number of antennas. The average CPU time requirements of different detectors under different scenarios are also provided for a more intuitive comparison, which show a consistent trend as our analysis. In particular, the low-complexity design LCRADMMNet requires nearly an order of magnitude less CPU time than RADMMNet. Concerning the SER performances presented before, although the non-robust detection networks may require less complexity than the proposed networks, they suffer from serious performance degradation. Meanwhile, it can be found that RADMMNet can achieve comparable performance to the optimal robust ML detection within only a few layers, while LCRADMMNet can always outperform ROAMPNet with much less complexity. Hence, we conclude that both the proposed networks achieve attractive tradeoffs between the performance and the complexity.} Regarding the considered two channel acquisition methods, the complexity of LMMSE channel estimation is ${\cal O}\left({M^3}{K^3}\right)$, while the proposed RDAKF-based channel tracking method with much better NMSE performance additionally requires a complexity of ${\cal O}\left(M^3L\right)$ per data block, due to the matrix inversion involved in the calculation of the Kalman gain. %

	\section{Conclusion}
This paper investigated the statistically robust detector design for MIMO systems by taking into account both channel estimation error and channel variation. We first derived an ADMM-based robust detection algorithm, which admits the calculations of closed-form expressions in each iteration. {Then, by deep unfolding the ADMM iterations and introducing some data-driven structures, we advocated a robust detection network RADMMNet with trainable parameters.} Furthermore, by adopting a Gaussian approximation for  the CSI error, a low-complexity robust MIMO detector was further developed along with the corresponding deep unfolding network LCRADMMNet. In addition, as a complementary way to combat the channel variation and enhance the detection performance, we also presented a Kalman-filter-based channel tracking method by fully exploiting the pre-estimated data symbols. Simulation results confirmed that, the proposed two networks can considerably outperform the non-robust detectors and even approach the optimal robust ML detector with much lower complexities under both the LMMSE channel estimation and the proposed RDAKF-based channel tracking method.
%validating the effectiveness of the derived ADMM-based robust detector and the superiority of the well-trained parameters.   
\end{sloppypar}

	%\section{}

	%\section*{Acknowledgment}

	\bibliography{RobustADMM}

\begin{thebibliography}{10}
\providecommand{\url}[1]{#1}
\csname url@samestyle\endcsname
\providecommand{\newblock}{\relax}
\providecommand{\bibinfo}[2]{#2}
\providecommand{\BIBentrySTDinterwordspacing}{\spaceskip=0pt\relax}
\providecommand{\BIBentryALTinterwordstretchfactor}{4}
\providecommand{\BIBentryALTinterwordspacing}{\spaceskip=\fontdimen2\font plus
\BIBentryALTinterwordstretchfactor\fontdimen3\font minus
  \fontdimen4\font\relax}
\providecommand{\BIBforeignlanguage}[2]{{%
\expandafter\ifx\csname l@#1\endcsname\relax
\typeout{** WARNING: IEEEtran.bst: No hyphenation pattern has been}%
\typeout{** loaded for the language `#1'. Using the pattern for}%
\typeout{** the default language instead.}%
\else
\language=\csname l@#1\endcsname
\fi
#2}}
\providecommand{\BIBdecl}{\relax}
\BIBdecl

\bibitem{conf}
Y.~Sun, H.~Shen, W.~Xu, and C.~Zhao, ``Learning statistically robust {MIMO}
  detection with imperfect {CSI},'' in \emph{Proc. 18th Int. Symp. Wireless
  Commun. Syst. (ISWCS)}, Hangzhou, China, Oct. 2022, pp. 1--6.

\bibitem{MIMO8944280}
M.~A. Albreem, M.~Juntti, and S.~Shahabuddin, ``Massive {MIMO} detection
  techniques: A survey,'' \emph{IEEE Commun. Surveys Tuts.}, vol.~21, no.~4,
  pp. 3109--3132, 4th Quart. 2019.

\bibitem{50YEAR}
S.~Yang and L.~Hanzo, ``Fifty years of {MIMO} detection: The road to
  large-scale {MIMOs},'' \emph{IEEE Commun. Surveys Tuts.}, vol.~17, no.~4, pp.
  1941--1988, 4th Quart. 2015.

\bibitem{LTE}
M.~Wu, B.~Yin, G.~Wang, C.~Dick, J.~R. Cavallaro, and C.~Studer, ``Large-scale
  {MIMO} detection for {3GPP} {LTE}: Algorithms and {FPGA} implementations,''
  \emph{IEEE J. Sel. Topics Signal Process.}, vol.~8, no.~5, pp. 916--929, Oct.
  2014.

\bibitem{SD}
M.~O. Damen, H.~El~Gamal, and G.~Caire, ``On maximum-likelihood detection and
  the search for the closest lattice point,'' \emph{IEEE Trans. Inf. Theory},
  vol.~49, no.~10, pp. 2389--2402, Oct. 2003.

\bibitem{EP}
T.~P. Minka, ``A family of algorithms for approximate {Bayesian} inference,''
  Ph.D. dissertation, MIT, Cambridge, MA, 2001.

\bibitem{AMP}
S.~Wu \emph{et~al.}, ``Low-complexity iterative detection for large-scale
  multiuser {MIMO-OFDM} systems using approximate message passing,'' \emph{IEEE
  J. Sel. Topics Signal Process.}, vol.~8, no.~5, pp. 902--915, Oct. 2014.

\bibitem{ADMM8933411}
S.~{Boyd}, N.~{Parikh}, E.~{Chu}, B.~{Peleato}, and J.~{Eckstein},
  ``Distributed optimization and statistical learning via the alternating
  direction method of multipliers,'' \emph{Found. Trends Mach. Learn.}, vol.~3,
  no.~1, pp. 1--122, Jan. 2011.

\bibitem{SNR2005}
T.~Weber, A.~Sklavos, and M.~Meurer, ``Imperfect channel-state information in
  {MIMO} transmission,'' \emph{IEEE Trans. Commun.}, vol.~54, no.~3, pp.
  543--552, Mar. 2006.

\bibitem{Statis8944280}
X.~Zhang, D.~P. Palomar, and B.~Ottersten, ``Statistically robust design of
  linear {MIMO} transceivers,'' \emph{IEEE Trans. Signal Process.}, vol.~56,
  no.~8, pp. 3678--3689, Aug. 2008.

\bibitem{Transceiver}
M.~Ding and S.~D. Blostein, ``{MIMO} minimum total {MSE} transceiver design
  with imperfect {CSI} at both ends,'' \emph{IEEE Trans. Signal Process.},
  vol.~57, no.~3, pp. 1141--1150, Oct. 2009.

\bibitem{Rtransceiver}
N.~Vucic, H.~Boche, and S.~Shi, ``Robust transceiver optimization in downlink
  multiuser {MIMO} systems,'' \emph{IEEE Trans. Signal Process.}, vol.~57,
  no.~9, pp. 3576--3587, Sep. 2009.

\bibitem{WidelyLinear}
Y.~Chen, L.~You, A.-A. Lu, and X.~Gao, ``Widely-linear processing for the
  uplink of the massive {MIMO} with {IQ} imbalance: Channel estimation and data
  detection,'' \emph{IEEE Trans. Signal Process.}, vol.~69, pp. 4685--4698,
  Aug. 2021.

\bibitem{MEP}
J.~Zeng, T.~Lv, R.~P. Liu, X.~Su, N.~C. Beaulieu, and Y.~J. Gu, ``Linear
  minimum error probability detection for massive {MU-MIMO} with imperfect
  {CSI} in {URLLC},'' \emph{IEEE Commun. Surveys Tuts.}, vol.~68, no.~11, pp.
  11384--11388, Oct. 2019.

\bibitem{Model8509622}
H.~He, C.~K. Wen, S.~Jin, and G.~Y. Li, ``Model-driven deep learning for {MIMO}
  detection,'' \emph{IEEE Trans. Signal Process.}, vol.~68, pp. 1702--1715,
  Feb. 2020.

\bibitem{OAMPNet}
H.~He, C.~K. Wen, S.~Jin, and G.~Y. Li, ``A model-driven deep learning network
  for {MIMO} detection,'' in \emph{Proc. IEEE Global Conf. Signal Inf. Process.
  (GlobalSIP)}, Anaheim, CA, USA, Nov. 2018, pp. 584--588.

\bibitem{Any2005}
P.~Garg, R.~K. Mallik, and H.~M. Gupta, ``Optimum receiver design and
  performance analysis of arbitrarily correlated {Rician} fading {MIMO}
  channels with imperfect channel state information,'' \emph{IEEE Trans. Inf.
  Theory}, vol.~56, no.~3, pp. 1114--1134, Mar. 2010.

\bibitem{Data7894280}
J.~Choi, ``Data detection with imperfect {CSI} using averaged likelihood
  function,'' \emph{IEEE Trans. Wireless Commun.}, vol.~7, no.~11, pp.
  4117--4121, Nov. 2008.

\bibitem{QAM86}
Y.~V. Zakharov, V.~M. Baronkin, and J.~Zhang, ``Optimal and mismatched
  detection of {QAM} signals in fast fading channels with imperfect channel
  estimation,'' \emph{IEEE Trans. Wireless Commun.}, vol.~8, no.~2, pp.
  617--621, Feb. 2009.

\bibitem{reduce}
B.~S. Thian and A.~Goldsmith, ``Reduced-complexity robust {MIMO} decoders,''
  \emph{IEEE Trans. Wireless Commun.}, vol.~12, no.~8, pp. 3783--3795, Aug.
  2013.

\bibitem{Deep8663966}
Z.~{Qin}, H.~{Ye}, G.~Y. {Li}, and B.-H. {Juang}, ``Deep learning in physical
  layer communications,'' \emph{IEEE Wireless Commun.}, vol.~26, no.~2, pp.
  93--99, Apr. 2019.

\bibitem{Model5338}
H.~{He}, S.~{Jin}, C.~{Wen}, F.~{Gao}, G.~Y. {Li}, and Z.~{Xu}, ``Model-driven
  deep learning for physical layer communications,'' \emph{IEEE Wireless
  Commun.}, vol.~26, no.~5, pp. 77--83, Oct. 2019.

\bibitem{unroll}
V.~Monga, Y.~Li, and Y.~C. Eldar, ``Algorithm unrolling: Interpretable,
  efficient deep learning for signal and image processing,'' \emph{IEEE Signal
  Process. Mag.}, vol.~38, no.~2, pp. 18--44, Mar. 2021.

\bibitem{survey}
M.~A. Albreem, A.~H. Alhabbash, S.~Shahabuddin, and M.~Juntti, ``Deep learning
  for massive {MIMO} uplink detectors,'' \emph{IEEE Commun. Surveys Tuts.},
  vol.~24, no.~1, pp. 741--766, 1st Quart. 2022.

\bibitem{xw2}
W.~Xu, Z.~Yang, D.~W.~K. Ng, M.~Levorato, Y.~C. Eldar, and M.~Debbah, ``Edge
  learning for {B5G} networks with distributed signal processing: Semantic
  communication, edge computing, and wireless sensing,'' \emph{IEEE J. Sel.
  Topics Signal Process.}, vol.~17, no.~1, pp. 9--39, Jan. 2023.

\bibitem{chanest}
M.~{Soltani}, V.~{Pourahmadi}, A.~{Mirzaei}, and H.~{Sheikhzadeh}, ``Deep
  learning-based channel estimation,'' \emph{IEEE Commun. Lett.}, vol.~23,
  no.~4, pp. 652--655, Apr. 2019.

\bibitem{ICINet}
Y.~Sun, H.~Shen, Z.~Du, L.~Peng, and C.~Zhao, ``{ICINet}: {ICI}-aware neural
  network based channel estimation for rapidly time-varying {OFDM} systems,''
  \emph{IEEE Commun. Lett.}, vol.~25, no.~9, pp. 2973--2977, Jun. 2021.

\bibitem{Precoder}
Q.~Hu, Y.~Cai, Q.~Shi, K.~Xu, G.~Yu, and Z.~Ding, ``Iterative algorithm induced
  deep-unfolding neural networks: Precoding design for multiuser {MIMO}
  systems,'' \emph{IEEE Trans. Wireless Commun.}, vol.~20, no.~2, pp.
  1394--1410, Feb. 2021.

\bibitem{linqi}
Q.~Lin, H.~Shen, and C.~Zhao, ``Learning linear {MMSE} precoder for uplink
  massive {MIMO} systems with one-bit {ADCs},'' \emph{IEEE Wireless Commun.
  Lett.}, vol.~11, no.~10, pp. 2235--2239, Oct. 2022.

\bibitem{DetNet52521}
N.~Samuel, T.~Diskin, and A.~Wiesel, ``Learning to detect,'' \emph{IEEE Trans.
  Signal Process.}, vol.~67, no.~10, pp. 2554--2564, May 2019.

\bibitem{LAN8715338}
M.~Kim and D.~Park, ``Learnable {MIMO} detection networks based on inexact
  {ADMM},'' \emph{IEEE Trans. Wireless Commun.}, vol.~20, no.~1, pp. 565--576,
  Jan. 2021.

\bibitem{LCG}
Y.~Wei, M.-M. Zhao, M.~Hong, M.-J. Zhao, and M.~Lei, ``Learned conjugate
  gradient descent network for massive {MIMO} detection,'' \emph{IEEE Trans.
  Signal Process.}, vol.~68, pp. 6336--6349, Jan. 2020.

\bibitem{Binary}
M.~Shao and W.-K. Ma, ``Binary {MIMO} detection via homotopy optimization and
  its deep adaptation,'' \emph{IEEE Trans. Signal Process.}, vol.~69, pp.
  781--796, Feb. 2021.

\bibitem{Low1312}
I.~N. Tiba, Q.~Zhang, J.~Jiang, and Y.~Wang, ``A low-complexity {ADMM}-based
  massive {MIMO} detectors via deep neural networks,'' in \emph{Proc. IEEE Int.
  Conf. Acoust., Speech Signal Process. (ICASSP)}, Toronto, ON, Canada, Jun.
  2021, pp. 1--5.

\bibitem{xw1}
H.~Huo, J.~Xu, G.~Su, W.~Xu, and N.~Wang, ``Intelligent {MIMO} detection using
  meta learning,'' \emph{IEEE Wireless Commun. Lett.}, vol.~11, no.~10, pp.
  2205--2209, Oct. 2022.

\bibitem{power}
H.~{Ye}, G.~Y. {Li}, and B.-H. {Juang}, ``Power of deep learning for channel
  estimation and signal detection in {OFDM} systems,'' \emph{IEEE Wireless
  Commun. Lett.}, vol.~7, no.~1, pp. 114--117, Feb. 2018.

\bibitem{ADMIN}
S.~Shahabuddin, I.~Hautala, M.~Juntti, and C.~Studer, ``{ADMM}-based
  infinity-norm detection for massive {MIMO}: Algorithm and {VLSI}
  architecture,'' \emph{IEEE Trans. VLSI Syst.}, vol.~29, no.~4, pp. 747--759,
  Apr. 2021.

\bibitem{Efficient75673}
Q.~Zhang, J.~Wang, and Y.~Wang, ``Efficient {QAM} signal detector for massive
  {MIMO} systems via {PS/DPS-ADMM} approaches,'' \emph{IEEE Trans. Wireless
  Commun.}, vol.~21, no.~10, pp. 8859--8871, May 2022.

\bibitem{DACE}
J.~Ma and L.~Ping, ``Data-aided channel estimation in large antenna systems,''
  \emph{IEEE Trans. Signal Process.}, vol.~62, no.~12, pp. 3111--3124, Jun.
  2014.

\bibitem{Markov}
A.~Lu, X.~Gao, W.~Zhong, C.~Xiao, and X.~Meng, ``Robust transmission for
  massive {MIMO} downlink with imperfect {CSI},'' \emph{IEEE Trans. Commun.},
  vol.~67, no.~8, pp. 5362--5376, Apr. 2019.

\bibitem{9445013}
S.~Srivastava, C.~S.~K. Patro, A.~K. Jagannatham, and L.~Hanzo, ``Sparse,
  group-sparse, and online bayesian learning aided channel estimation for
  doubly-selective {mmWave} hybrid {MIMO OFDM} systems,'' \emph{IEEE Trans.
  Commun.}, vol.~69, no.~9, pp. 5843--5858, Jun. 2021.

\bibitem{zhangyu}
Y.~Zhang, D.~Wang, Y.~Huo, X.~Dong, and X.~You, ``Hybrid beamforming design for
  {mmWave} {OFDM} distributed antenna systems,'' \emph{Sci. China Inf. Sci.},
  vol.~63, no.~9, pp. 221--232, Jul. 2020.

\bibitem{CG123}
B.~Yin, M.~Wu, J.~R. Cavallaro, and C.~Studer, ``Conjugate gradient based
  soft-output detection and precoding in massive {MIMO} systems,'' in
  \emph{Proc. IEEE Global Commun. Conf. (GLOBECOM)}, Austin, TX, USA, Dec.
  2014, pp. 3696--3701.

\bibitem{ParasADMM}
E.~Ghadimi, A.~Teixeira, I.~Shames, and M.~Johansson, ``Optimal parameter
  selection for the alternating direction method of multipliers ({ADMM}):
  Quadratic problems,'' \emph{IEEE Trans. Autom. Control}, vol.~60, no.~3, pp.
  644--658, Mar. 2015.

\bibitem{Kron5338}
S.~L. Loyka, ``Channel capacity of {MIMO} architecture using the exponential
  correlation matrix,'' \emph{IEEE Commun. Lett.}, vol.~5, no.~9, pp. 369--371,
  Sep. 2001.

\end{thebibliography}
	
\end{document}